\documentclass[aps,twocolumn,prd,nofootinbib,longbibliography,superscriptaddress,floatfix]{revtex4-2}

\usepackage{amsfonts}
\usepackage{ulem}
\usepackage{tikz}
\usepackage{mathrsfs}
\usepackage{graphicx}
\usepackage{dcolumn}
\usepackage{bm}
\usepackage{bbm}
\usepackage[unicode, colorlinks, allcolors=blue!70!black, linktocpage, pdfusetitle]{hyperref}

\usepackage{color}
\usepackage{braket}
\usepackage{ulem}
\usepackage[all]{hypcap}
\definecolor{indigo(dye)}{rgb}{0.0, 0.25, 0.42}
\usepackage{amsthm}

\theoremstyle{definition}

\newif\ifslow

\tikzset{>=latex} 
\colorlet{myred}{red!80!black}
\colorlet{myblue}{blue!80!black}
\colorlet{mygreen}{green!80!black}
\colorlet{mydarkred}{red!50!black}
\colorlet{mydarkblue}{blue!50!black}
\colorlet{mylightblue}{mydarkblue!3}
\colorlet{mypurple}{blue!40!red!80!black}
\colorlet{mydarkpurple}{blue!40!red!50!black}
\colorlet{mylightpurple}{mydarkpurple!40!red!3}
\colorlet{myorange}{orange!40!yellow!95!black}
\tikzstyle{cone}=[mydarkblue,line width=0.2,top color=blue!60!black!30,
                  bottom color=blue!60!black!50!red!30,shading angle=60,fill opacity=0.9]
\tikzstyle{cone back}=[mydarkblue,line width=0.1,dash pattern=on 1pt off 1pt]
\tikzstyle{photon}=[-{Latex[length=4,width=3]},myorange,line width=0.4,decorate,
                    decoration={snake,amplitude=0.9,segment length=4,post length=3.8}]
\tikzstyle{world line}=[myblue!30,line width=0.4]
\tikzstyle{world line t}=[mypurple!30,line width=0.4]
\tikzstyle{particle}=[mygreen,line width=0.5]
\tikzstyle{singularity}=[line width=0.6,decorate,
                         decoration={zigzag,amplitude=2,segment length=6.17}]
                         
\tikzset{declare function={%
  penrose(\x,\c)  = {\fpeval{2/pi*atan( (sqrt((1+tan(\x)^2)^2+4*\c*\c*tan(\x)^2)-1-tan(\x)^2) /(2*\c*tan(\x)^2) )}};%
  penroseu(\x,\t) = {\fpeval{atan(\x+\t)/pi+atan(\x-\t)/pi}};%
  penrosev(\x,\t) = {\fpeval{atan(\x+\t)/pi-atan(\x-\t)/pi}};%
  kruskal(\x,\c)  = {\fpeval{asin( \c*sin(2*\x) )*2/pi}};
}}

\def\R{0.08} 

\usepackage{xfp} 
\usepackage[outline]{contour} 
\usetikzlibrary{decorations.markings,decorations.pathmorphing}
\usetikzlibrary{angles,quotes} 
\usetikzlibrary{arrows.meta} 
\contourlength{1.4pt}

\newcommand{\defn}{\mathrel{\mathop:}=} 

\newcommand{\be}{\begin{equation}}
\newcommand{\ee}{\end{equation}}

\newcommand{\mc}{\mathcal}

\newcommand{\ms}{\mathscr}

\usepackage{yhmath}

\newcommand{\Lie}{\pounds}
\newcommand{\hatLie}{\Lie\kern-0.25em\hat{\vphantom{\Lie{}}}\kern0.25em}

\begin{document}
\count\footins = 1000

\title{How to Minimize the Decoherence Caused by Black Holes}

\author{Daine L. Danielson}
\email{daine@uchicago.edu}
\affiliation{Enrico Fermi Institute, Kadanoff Center for Theoretical Physics, and Department of Physics, The University of Chicago, Chicago, IL 60637, USA}
\author{Jonah Kudler-Flam}%
 \email{jkudlerflam@ias.edu}
  \affiliation{School of Natural Sciences, Institute for Advanced Study, Princeton, NJ 08540, USA}
 \affiliation{Princeton Center for Theoretical Science, Princeton University, Princeton, NJ 08544, USA}
\author{Gautam Satishchandran}%
 \email{gautam.satish@princeton.edu}
 \affiliation{Princeton Gravity Initiative, Princeton University, Princeton, NJ 08544, USA}
 \author{Robert M. Wald}
\email{rmwa@uchicago.edu }
\affiliation{Enrico Fermi Institute, Kadanoff Center for Theoretical Physics, and Department of Physics, The University of Chicago, Chicago, IL 60637, USA}

\begin{abstract}
\noindent We consider an experimentalist, Alice, who creates a quantum superposition of a charged or massive body outside of a black hole (or, more generally, in the presence of a Killing horizon). It was previously shown that emission of soft photons/gravitons into the black hole will result in the decoherence of the components of the superposition if it is held open for a sufficiently long span of time. However, at any finite time, $t_c$, during the process, it is not obvious how much decoherence has irrevocably occurred. Equivalently, it is not obvious how much information an observer inside the black hole can extract about Alice's superposition prior to time $t_c$. In this paper, we solve for the optimal experimental protocol to be followed by Alice for $t > t_c$ so as to minimize the decoherence of the components of her superposition. More precisely, given the entangling radiation that has passed through the horizon prior to the cross-section $\mathcal C$ corresponding to the time $t = t_c$ in Alice's lab, we determine the ``optimal purification'' of this radiation beyond $\mathcal C$ such that the global quantum state of the radiation through the horizon has maximal overlap (quantum fidelity) with the Hartle-Hawking or Unruh vacuum. Due to the intricate low frequency entanglement structure of the quantum field theory vacuum state, we find this optimal purification to be nontrivial. In particular, even if Alice has already ``closed'' her superposition by bringing the components back together, we find that she can decrease the net decoherence of the components of her superposition somewhat by reopening it and performing further manipulations.
\end{abstract}

\maketitle

\section{Introduction}
In any quantum mechanical experiment or computation, some degree of quantum entanglement is inevitably generated between the initially isolated quantum mechanical degrees of freedom and external degrees of freedom, resulting in some degree of decoherence of the body.
An interesting example of the decoherence of a quantum body arises when the body is coupled to a long-range quantum field, such as a charged body coupled to a quantum electromagnetic field or a massive body coupled to a quantum gravitational field (which we will treat as a linearized quantum field in a background classical spacetime). The long-range field may mediate interactions of the body under consideration with external quantum systems, resulting in entanglement with these systems and a corresponding ``environmental decoherence'' of the body. However, even if no other systems are present, decoherence may result from the entanglement of the body with radiation emitted by the body. In this paper, we will be concerned with the decoherence resulting from emission of radiation.

Hereinafter, we will refer to the quantum body under consideration as a ``particle,'' but we primarily have in mind a ``nanoparticle'' rather than an elementary particle.
 If one puts the particle into a superposition of spatially separated states---e.g., by putting the particle through a Stern-Gerlach apparatus---the long-range quantum field will contain ``information'' about the superposition that could, in principle, be harvested by a distant observer if given sufficient time\footnote{{Quantum spatial superpositions of massive bodies have been of recent interest in both theoretical as well as proposed experimental probes of fundamental properties of quantum gravity, e.g., \cite{Bose_2017,Marletto_2017,Belenchia_2018,Christodoulou_2019,Giacomini_2019,Aspelmeyer_2021,Danielson:2021egj,Carney_2021,christodoulou_2022,carney_2022,Feng_2022,Zhou_2022,Overstreet_2022}.}}. One might think that the distinguishability of the quantum field associated with the different components of the superposition would give rise to the decoherence of the superposition. However, this is not necessarily the case. If there is no external system present to harvest the information and if no radiation is emitted, then any apparent decoherence associated with the distinguishable long-range fields would, in fact, constitute ``false decoherence'' \cite{Unruh_2000}. 
 In particular, in empty flat spacetime, if the superposition is opened and closed in a sufficiently adiabatic manner so that negligible radiation is emitted, then the quantum field states associated with the components of the superposition will return to their initial values, and no decoherence will result. 

Although the decoherence of such a superposition can be made negligible in Minkowski spacetime, it was recently shown that the situation changes dramatically if the particle is in the exterior of a black hole, or more generally in the vicinity of a Killing horizon \cite{2022IJMPD..3141003D,Danielson2022sga} (see also \cite{Gralla:2023oya}, \cite{Li_2024}). To see this more concretely, consider an experimentalist, Alice, who controls a quantum particle. Suppose that the particle has spin initially in the $x$ direction. Alice then puts the particle through a $z$-Stern-Gerlach apparatus. She then keeps the components of the particle spatially separated for a time $T$, after which she recombines the components of the particle's wave function via a reversing Stern-Gerlach apparatus. At the end of the process, the state of the joint particle-and-field system will be given by \cite{Danielson2022sga}
\begin{align}
    \frac{1}{\sqrt{2}}\left( \ket{\uparrow }\otimes \ket{\psi_1} + \ket{\downarrow}\otimes \ket{\psi_2}\right),
\end{align}
where $\ket{\psi_1}$ and $\ket{\psi_2}$ are the coherent radiation states
of the quantum field sourced, respectively, by the spin-up and spin-down components of the particle. The density matrix of Alice's particle is
\begin{align}
    \rho = \frac{1}{2}\begin{pmatrix}
        1 & \bra{\psi_2}\psi_1\rangle
        \\
        \bra{\psi_1}\psi_2\rangle & 1
    \end{pmatrix}.
\end{align}
The decoherence, $\ms{D}$, of the spin-up and spin-down components is controlled by the size of the off-diagonal elements, i.e., the overlap of states of the radiation emitted by the spin-up and spin-down components: 
\begin{align}
    \ms{D} \defn 1 - |\langle\psi_1|\psi_2\rangle|. 
\end{align}
In Minkowski spacetime, if the separation and recombination are performed sufficiently adiabatically, we will have $\ket{\psi_1} \approx \ket{\psi_2} \approx \ket{0}$ and we find $\ms{D} \approx 0$ \cite{Belenchia_2018,dswLocal_2024}. In particular, the length of time, $T$, over which the particle components are kept separated does not contribute to the decoherence. However, in the presence of a black hole, the radiation states $\ket{\psi_1}$ and $\ket{\psi_2}$ include the radiation that goes into the black hole, and ``soft radiation''\footnote{{This soft radiation is due to a ``horizon memory effect'' \cite{Hawking:2016msc,Donnay:2018ckb,Rahman:2019bmk,2022IJMPD..3141003D,Danielson2022sga} which is an exact mathematical analog of the soft radiation and associated memory effect at null infinity in asymptotically flat spacetimes (see, e.g., \cite{asymp-quant,Strominger:2017zoo,Satishchandran:2019pyc,Prabhu:2022zcr,Prabhu:2024zwl,Prabhu:2024lmg}).}} will unavoidably be emitted into the black hole. For large $T$, the number of photons/gravitons in this soft radiation grows linearly with $T$ \cite{2022IJMPD..3141003D,Danielson2022sga,dswLocal_2024}, resulting in a decoherence that grows with $T$ as 
\begin{align}
    \ms{D}  \simeq  1- e^{-\Gamma_{{\rm BH}} T}
\end{align}
where $\Gamma_{{\rm BH}}$ depends upon the field under consideration as well as the various details of the process. This effect can also be seen to equivalently arise from local interactions of Alice's particle with the vacuum in the black hole exterior \cite{dswLocal_2024,Wilson-Gerow:2024ljx}. This formulation can be directly compared to the decohering effects arising from entanglement of the particle with any internal degrees of freedom of a material body \cite{Biggs:2024dgp,dswLocal_2024}. Of course, $\Gamma_{{\rm BH}}$ will be very small unless the process is done near the black hole, with large charges/masses and large separations. Expressions for $\Gamma_{{\rm BH}}$ can be found in \cite{2022IJMPD..3141003D,Gralla:2023oya}.

Although the analysis of \cite{2022IJMPD..3141003D,Danielson2022sga} gives the decoherence in the presence of a black hole that results from the complete process of separation and recombination of the components of the superposition, the question remains as to exactly ``when'' the decoherence is occurring. Specifically, consider a time $t= t_c$ in Alice's lab after the separation of the components has taken place but before the recombination has been completed. How much decoherence has irrevocably occurred by this time? Equivalently, in principle, how much ``which path'' information about Alice's superposition can an observer inside the black hole gather prior to entering the future of Alice's lab at time $t=t_c$? 

The main purpose of this paper is to answer this question. We will do so by supposing that Alice has followed some fixed protocol for her experiment for times $t \leq t_c$. Let $\mathcal C$ denote the cross-section of the black hole horizon that corresponds to the time $t=t_c$ in Alice's laboratory; i.e., $\mathcal C$ bounds the portion of the horizon that lies outside of the future of Alice's laboratory at time $t=t_c$. The protocol that Alice follows up to time $t=t_c$ in her laboratory will result in some given state of entangling radiation that crosses the event horizon prior to $\mathcal C$. We shall determine the ``optimal purification'' of this radiation state, i.e., the state of radiation on the horizon to the future of $\mathcal C$ that maximizes $|\langle{\psi_1}|\psi_2\rangle|$. This quantity, for the optimal purification, is sometimes referred to as the (Uhlmann) fidelity \cite{uhlmann1976transition}. The decoherence obtained by using this optimal purification will be interpreted as the decoherence that has irrevocably occurred for Alice's particle by time $t_c$.

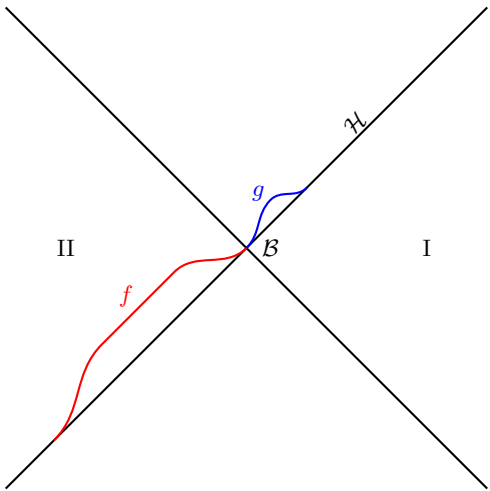
\begin{figure}
    \centering
\begin{tikzpicture}[scale=3.2]
  \message{Extended Penrose diagram: Schwarzschild black hole^^J}
  
  \def\R{0.08} 
  \def\Nlines{1} 
  \pgfmathsetmacro\ta{1/sin(90*1/(\Nlines+1))} 
  \pgfmathsetmacro\tb{sin(90*2/(\Nlines+1))}   
  \pgfmathsetmacro\tc{1/sin(90*2/(\Nlines+1))} 
  \pgfmathsetmacro\td{sin(90*1/(\Nlines+1))}   
  \coordinate (-O) at (-1, 0); 
  \coordinate (-S) at (-1,-1); 
  \coordinate (-N) at (-1, 1); 
  \coordinate (-W) at (-2, 0); 
  \coordinate (-E) at ( 0, 0); 
  \coordinate (O)  at ( 1, 0); 
  \coordinate (S)  at ( 1,-1); 
  \coordinate (N)  at ( 1, 1); 
  \coordinate (E)  at ( 2, 0); 
  \coordinate (W)  at ( 0, 0); 
  \coordinate (B)  at ( 0,-1); 
  \coordinate (T)  at ( 0, 1); 
  \coordinate (X0) at ({asin(sqrt((\ta^2-1)/(\ta^2-\tb^2)))/90},
                       {-acos(\ta*sqrt((1-\tb^2)/(\ta^2-\tb^2)))/90}); 
  \coordinate (X1) at ({asin(sqrt((\tc^2-1)/(\tc^2-\td^2)))/90},
                       {acos(\tc*sqrt((1-\td^2)/(\tc^2-\td^2)))/90}); 
  \coordinate (X2) at (45:0.87); 
  \coordinate (X3) at (0.60,1.05); 

  \draw[thick,black] (N) -- (-S) ;
  \draw[thick,black] (-N) -- (S) ;
  \node[] at (-.75,0) {II};
  \node[] at (.75,0) {I};

  \node[above=-2.5,rotate=45] at (.5,0.5) {$\mathcal{H}$};
  \node[] at (0.1,0) {$\mathcal{B}$};
  \draw[red,thick] (-.8,-.8) to[out = 45, in = -135] (-.6,-.4) to[out = 45, in = -135] (-.3,-.1) to[out = 45, in = -135] (0,0);
  \draw[blue, thick](0,0) to[out = 45, in = -135]  (.1,.2)to[out = 45, in = -135] (.25,.25);
  \node[red] at (-.5,-.2) {$f$};
  \node[blue] at (.05,.23) {$g$};
\end{tikzpicture}
    \caption{A spacetime diagram illustrating a Rindler horizon in Minkowski spacetime, which consists of two null hyperplanes that intersect at the bifurcation surface $\mathcal B$. The null hyperplane labeled $\mathcal H$ can be treated as an initial data surface for solutions that fall off at infinity. The red curve shows the graph of the restriction to $\mathcal{H}$ of a solution $f$ with initial data with support in Rindler wedge II, whereas the blue curve shows the restriction to $\mathcal{H}$ of a solution $g$ with initial data with support in wedge I.}
    \label{fig:bifurcate_tikz}
\end{figure}

We will find the optimal purification by first solving the following mathematical problem for Rindler wedges in Minkowski spacetime (see Fig.~\ref{fig:bifurcate_tikz}).
Suppose we are given a classical solution $f$ with initial data with support in Rindler wedge II; i.e., $f$ vanishes in Rindler wedge I. The corresponding quantum coherent state associated with $f$ will then coincide with the Minkowski vacuum in wedge I. We wish to determine the classical solution, $g$, with initial data with support in Rindler wedge I that has the property that the coherent state associated with $f +g$ has maximal overlap with the Minkowski vacuum. It is not difficult to see that this will be the case if and only if $g$ is such that the Klein-Gordon norm of the positive frequency part (with respect to inertial time) of $f+g$ is minimized.

In fact, it is convenient to reformulate this question as follows: Let $\mathcal H$ denote one of the two null planes that comprise the Rindler horizon, as illustrated in Fig.~\ref{fig:bifurcate_tikz}. As argued in \cite{unruh1984happens}, $\mathcal H$ is an appropriate initial data surface for solutions that fall off at infinity, so we can characterize $f$ and $g$ by their restrictions to $\mathcal H$. On $\mathcal H$, $f$ will have support to the past of the bifurcation surface, $\mathcal B$, of the horizon, whereas any solution $g$ with support in Rindler wedge I will have data on $\mathcal H$ with support to the future of $\mathcal B$ as shown in Fig.~\ref{fig:bifurcate_tikz}. We wish to find the function $g$ with support to the future of $\mathcal B$ that minimizes the Klein-Gordon norm on $\mathcal H$ of the positive frequency part of $f+g$ with respect to affine time on $\mathcal H$ (corresponding to positive inertial frequency).

We will analyze this question in Sec.~\ref{rindler}. We will show there that the optimal $g$ is obtained by the following sequence of steps: (i) Reflect $f$ about the bifurcation surface $\mathcal B$ to get a function $\tilde{f}$ with support to the future of $\mathcal B$. (ii) Decompose $\tilde{f}$ into modes of definite Rindler (i.e., boost Killing) frequency. (iii) Multiply each mode by an appropriate thermal factor and resum to get $g$. We will show that steps (ii) and (iii) are equivalent to obtaining $g$ from $\tilde{f}$ via the formula 
\begin{align}
    g(t,y) = \kappa\int_{-\infty}^{\infty} \tilde{f}(t',y) {\text{sech}\left(\kappa  [t-t'] \right/2)} dt'
\label{gpres}
\end{align}
where $t$ denotes the boost Killing time on the portion of the horizon to the future of the bifurcation surface, 
and $\kappa$ is the surface gravity of the horizon. 

We will also need to solve the more general problem where $f$ is not required to vanish on the bifurcation surface, $\mathcal B$. We then seek the continuation, $g$, of $f$ to the future of $\mathcal B$ that minimizes the Klein-Gordon norm of the positive frequency part of $f+g$ with respect to affine parameter. We will argue at the end of Sec.~\ref{rindler} that Eq.~\eqref{gpres} remains the optimal choice of $g$ in this more general situation.

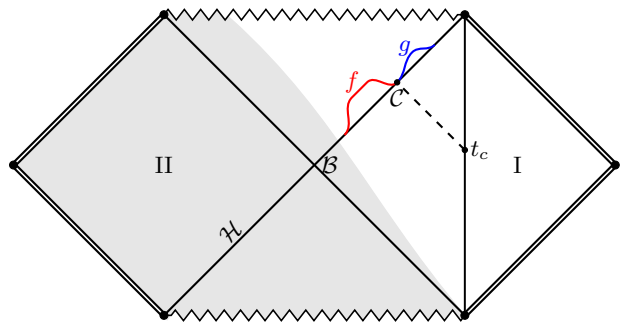
\begin{figure}
    \centering
\begin{tikzpicture}[scale=2]
  \message{Extended Penrose diagram: Schwarzschild black hole^^J}
  
  \def\R{0.08} 
  \def\Nlines{1} 
  \pgfmathsetmacro\ta{1/sin(90*1/(\Nlines+1))} 
  \pgfmathsetmacro\tb{sin(90*2/(\Nlines+1))}   
  \pgfmathsetmacro\tc{1/sin(90*2/(\Nlines+1))} 
  \pgfmathsetmacro\td{sin(90*1/(\Nlines+1))}   
  \coordinate (-O) at (-1, 0); 
  \coordinate (-S) at (-1,-1); 
  \coordinate (-N) at (-1, 1); 
  \coordinate (-W) at (-2, 0); 
  \coordinate (-E) at ( 0, 0); 
  \coordinate (O)  at ( 1, 0); 
  \coordinate (S)  at ( 1,-1); 
  \coordinate (N)  at ( 1, 1); 
  \coordinate (E)  at ( 2, 0); 
  \coordinate (W)  at ( 0, 0); 
  \coordinate (B)  at ( 0,-1); 
  \coordinate (T)  at ( 0, 1); 
  \coordinate (X0) at ({asin(sqrt((\ta^2-1)/(\ta^2-\tb^2)))/90},
                       {-acos(\ta*sqrt((1-\tb^2)/(\ta^2-\tb^2)))/90}); 
  \coordinate (X1) at ({asin(sqrt((\tc^2-1)/(\tc^2-\td^2)))/90},
                       {acos(\tc*sqrt((1-\td^2)/(\tc^2-\td^2)))/90}); 
  \coordinate (X2) at (45:0.87); 
  \coordinate (X3) at (0.60,1.05); 
  \definecolor{vlightgray}{rgb}{0.9,0.9,0.9}
  \begin{scope}
    
    \clip[decorate,decoration={zigzag,amplitude=2,segment length=6.17}]
      (S) -- (-S) --++ (-1.1,-0.1) |-++ (4.2,2.2) |- cycle;
    \clip[decorate,decoration={zigzag,amplitude=2,segment length=6.17}]
      (-N) -- (N) --++ (1.1,0.1) |-++ (-4.2,-2.2) |- cycle;

    \filldraw[vlightgray] (1.0,-1)to[out=130,in=-40] (-.7,1.1)-- (-1.1,1.1)--(-1,1)--(-2,0)--(-1,-1)--(-1,-1.1)--(1,-1.1)-- cycle;
  \end{scope}
  
  \draw[singularity] (-N) -- node[above] {} (N);
  \draw[singularity] (S) -- node[below] {} (-S);
  \path (S) -- (W) node[black,pos=0.50,below=-2.5,rotate=-45,scale=1] {};
  \path (W) -- (N) node[black,pos=-.5,above=-2.5,rotate=45,scale=1]
    {{ $\mathcal{H}$}};
  \draw[thick,black,double] (-N) -- (-W) -- (-S);
  \draw[thick,black] (-N) -- (-E) -- (-S);
   \path (-N) -- (-E) node[black,pos=0.50,below=-2.5,rotate=-45,scale=1] {};
       \path (-S) -- (-E) node[black,pos=0.50,below=-2.5,rotate=45,scale=1] {};
  \draw[thick,black,double] (N) -- (E) -- (S);
\draw[thick,black] (N) -- (W) -- (S);

  \node[inner sep=2] at (-1,0) {II};
  \node[inner sep=2] at (1.35,0) {I};

    \filldraw[black] (-W) circle (.75pt);
  \filldraw[black] (-N) circle (.75pt);
  \filldraw[black] (-S) circle (.75pt);
      \filldraw[black] (E) circle (.75pt);
  \filldraw[black] (N) circle (.75pt);
  \filldraw[black] (S) circle (.75pt);
  \node[] at (0.1,0) {$\mathcal{B}$};
    \draw[line width=0.75,postaction={decorate}]
      (S) to[out=90,in=-90](N);

    \draw[dashed, thick] (1,.1) -- (1-.45,.1+.45);
    \node[] at (1-.45,.1+.35) {$\mathcal{C}$};

    \node[] at (1.1,.1) {$t_c$};
    \filldraw[] (1,.1) circle (.5pt);
      \draw[red,thick] (-.35+.55,-.35+.55) to[out = 45, in = -135] (-.3+.55,-.1+.55) to[out = 45, in = -135] (-.2+.55,0+.55) to[out = 45, in = -135] (0+.55,0+.55);
  \draw[blue, thick](0+.55,0+.55) to[out = 45, in = -135]  (.1+.55,.2+.55)to[out = 45, in = -135] (.25+.55,.25+.55);
  \node[red] at (-.5+.75,-.2+.75) {$f$};
  \node[blue] at (.05+.55,.23+.55) {$g$};
      \filldraw[] (1-.45,.1+.45) circle (.5pt);
\end{tikzpicture}
    \caption{A spacetime diagram showing the exterior region of a black hole (region I), where Alice performs her experiment. The gray shaded region is shown only for comparison with an ``eternal black hole'' spacetime; for a physical black hole formed by gravitational collapse, the shaded region would be replaced by an interior solution for the collapsing matter. The time $t_c$ in Alice's lab corresponds (via null propagation) to the cross-section $\mathcal C$ on the event horizon $\mathcal H$. By the time $\mathcal C$, the entangling radiation $f$ has already passed through $\mathcal H$. We wish to find the ``optimal purification'' $g$ for times later than $\mathcal C$. 
    }
    \label{fig:BH_tikz}
\end{figure}

The relevance of the above mathematical problem to the problem of determining the optimal protocol for Alice to follow in her experiment outside of a black hole can be seen as follows. The event horizon of a stationary black hole is a Killing horizon, with properties very similar to that of $\mathcal H$. However, for a black hole formed by gravitational collapse that settles down to a stationary\footnote{{We do not consider times comparable to the evaporation timescale of the black hole.}} final state, the stationary portion of the horizon corresponds to a portion of $\mathcal H$ in Fig.~\ref{fig:bifurcate_tikz} that lies entirely to future of $\mathcal B$ (see Fig.~\ref{fig:BH_tikz}).
Thus, when Alice performs her experiment, the entangling radiation emitted for $t \leq t_c$ crosses the horizon to the past of some cross-section $\mathcal C$, which lies to the future of $\mathcal B$. The analysis given in Sec.~\ref{rindler} of the optimal purification for the Rindler case proceeds by the decomposing into modes of definite boost Killing frequency. In order to be able to do this, it was important that $f$ have support to the past of $\mathcal B$ and for $g$ to have support to the future of $\mathcal B$. Thus, it might appear that the optimal purification found in the Rindler case would be of little relevance for finding the optimal purification for an experiment performed outside of a stationary black hole formed by gravitational collapse.

However, the fact that $t$ is a Killing parameter of a true spacetime Killing field did not play an important role in the analysis. For the task at hand---namely, to find the function $g$ that minimizes the Klein-Gordon norm of the positive frequency part of $f+g$ with respect to affine parameter---there is no essential geometrical difference between $\mathcal C$ and $\mathcal B$. We can proceed as follows. On a bifurcate Killing horizon, the relation between affine time $V$ and Killing time $t$ to the future of the bifurcation surface $\mathcal B$ is given by
\begin{align}
    V = e^{\kappa t}
\end{align}
where we have chosen the origin of $V$ so that $V=0$ corresponds to $\mathcal B$. Let $\mathcal C$ be an arbitrary cross-section lying to the future of $\mathcal B$. Without loss of generality, we may rescale $V$ so that $\mathcal C$ is the surface $V = V_c$ (independent of $y$). To the future of $\mathcal C$, define the ``pseudo-Killing time'' $v$ by
\begin{align}
    V - V_c= e^{\kappa v}
\end{align}
Although a translation in $v$ does not correspond to an actual spacetime symmetry, the relevant mathematical properties of a decomposition into modes of definite $v$ frequency for $V > V_c$ are in exact correspondence with those of the decomposition into modes of definite $t$ frequency in the Rindler case. It follows immediately that the optimal purification in the black hole case is obtained by reflecting the radiation for $V < V_c$ about $\mathcal C$ and then applying Eq.~\eqref{gpres} with $v$ replacing $t$. We may then rewrite $g$ as a function of true Killing time $t$ using
\begin{align}
    v = \frac{1}{\kappa} \ln (V - V_c) = \frac{1}{\kappa} \ln (e^{\kappa t} - e^{\kappa t_c}) .
\end{align}

In Sec.~\ref{rindler}, we obtain the optimal purification in Minkowski spacetime of a coherent state in Rindler wedge I. In Sec.~\ref{bhmin}, we give a prescription that Alice must follow to minimize the decoherence of her superposition outside of a black hole or other Killing horizon. Some concluding remarks are given in Sec.~\ref{disc}.

\section{Optimal Purification in Rindler Spacetime}
\label{rindler}

For simplicity and definiteness, in this paper we will restrict consideration to a real, massless quantum scalar field obeying the Klein-Gordon equation
\begin{align}
    \nabla^a \nabla_a \Phi = 0.
\label{KG}
\end{align}
It {is} straightforward to generalize our results to a {massive scalar}, Maxwell field, or a linearized gravitational field. 

In this section, we will be concerned with the following problem: Let $f$ be a classical solution of Eq.~\eqref{KG} in $d$-dimensional Minkowski spacetime with initial data supported in Rindler wedge II (see Fig.~\ref{fig:bifurcate_tikz}). Consider the quantum coherent state on Minkowski spacetime associated with $f$. The restriction of this state to Rindler wedge II is a mixed state in that wedge. We seek a state on Minkowski spacetime that (i) is a ``purification'' of this state (i.e., it is a pure state on all of Minkowski spacetime such that its restriction to Rindler wedge II agrees with the original state) and (ii) is ``optimal'' in the sense that its overlap with the Minkowski vacuum is maximized. We will restrict our analysis of the optimal purification to coherent states, but we believe that the optimal coherent state that we will find will be optimal among all possible purifications. Since the overlap of a coherent state with the vacuum is determined by the norm of its one-particle component, the optimal coherent state will be obtained by finding a classical solution $g$ with initial data with support in Rindler wedge I such that the norm of the one-particle state corresponding to $f+g$ is minimized. Of course, one possible purification would be to simply take $g=0$, so that the state corresponds to the Minkowski vacuum in wedge I. However, we will see that this is not the optimal purification.

As already argued in the introduction, this problem can be reformulated as follows. Given a real valued function $f$ on the Rindler horizon $\mathcal H$ (see Fig.~\ref{fig:bifurcate_tikz}) with support to the past of the bifurcation surface $\mathcal B$, find the real valued function $g$ on $\mathcal H$ with support to the future of $\mathcal B$ such that the Klein-Gordon norm of the positive frequency part of $f+g$ with respect to affine time on $\mathcal H$ is minimized. Note that the requirement that $f$ have support on the past of $\mathcal B$ requires that $f$ vanishes on $\mathcal B$. This assumption will be used in our analysis below to ensure that the positive frequency part of $f$ with respect to Rindler time will have finite Klein-Gordon norm, so that our basis expansion formulas will be well defined. However, we will remove this restriction at the end of this section.

We take the boost Killing field to be future directed in wedge I and, thus, past directed in wedge II. Let $V$ denote an affine parameter on $\mathcal H$, with $V=0$ corresponding to $\mathcal B$. Let ${\mathcal H}_I$ denote the portion of $\mathcal H$ with $V > 0$, and let ${\mathcal H}_{II}$ denote the portion of $\mathcal H$ with $V < 0$. We will use the same letter $t$ to denote ``Rindler time'' (i.e., Killing parameter of the boost Killing field) on both ${\mathcal H}_{I}$ and ${\mathcal H}_{II}$. For $V > 0$, $t$ increases to the future, and the relationship between $t$ and $V$ is given by
\begin{align}
    V =  e^{\kappa t} \quad \quad V > 0
\label{ka1}
\end{align}
where $\kappa$ is the surface gravity of $\mathcal H$.
For $V < 0$, $t$ increases to the past and the relationship between $t$ and $V$ is given by
\begin{align}
    V = - e^{\kappa t} \quad \quad V < 0 \, .
\label{ka2}
\end{align} 

Since $\mathcal H$ is effectively a Cauchy surface for solutions that fall off at infinity \cite{unruh1984happens},
the Klein-Gordon inner product on $\mc{H}$ of two such solutions $\phi_1$ and $\phi_2$ is given by 
\begin{align}
    \langle \phi_1 , \phi_2\rangle_{{\rm KG}} = i\int_{\mathcal H} \left(\phi_1^* \partial_V \phi_2 -  \phi_2 \partial_V \phi{}^*_{1}\right) dV d^{d-2}y
\label{KGprod}
\end{align}
where $y$ denotes the transverse directions on $\mathcal H$. This inner product is positive definite when restricted to solutions that are positive frequency with respect to affine time $V$. It is also positive definite when restricted to solutions with support in ${\mathcal H}_I$ that are positive frequency with respect to Rindler time. On the other hand, on account of the reversal in the direction of Rindler time on ${\mathcal H}_{II}$, it is positive definite when restricted to solutions with support in ${\mathcal H}_{II}$ that are negative frequency with respect to Rindler time.

A key idea in our analysis is to decompose $f$ and $g$ into Fourier modes with respect to Rindler time $t$. However, modes that oscillate exactly as $e^{-i \omega t}$ are not normalizable and are singular at $\mathcal B$. For these reasons, it is preferable to decompose $f$ and $g$ into a basis of normalized wave packets that are very sharply peaked in Rindler frequency but fall off in Rindler time and, in particular, vanish at $\mathcal B$. A suitable orthonormal basis can be constructed as follows \cite{hawking1975particle,Wald_75}. First, let $\{ h_m(y) \}$ denote an $L^2$ orthonormal basis of functions in the transverse directions, $y$, on $\mathcal H$. Choose $\epsilon > 0$ and for all integers $j \geq 0$ and $n$, define,
\begin{align}
\label{eq:basis}
\phi_{{\rm I}jnm} = \frac{h_m(y)}{\sqrt{\epsilon}} \int_{j \epsilon}^{(j+1) \epsilon} \frac{e^{-2\pi i n \omega/\epsilon} }{\sqrt{4\pi\omega}} e^{-i \omega t} d \omega.
\end{align}
For notational simplicity, we will lump $n$ and $m$ into a single discrete index $k$ and denote $\phi_{Ijnm}$ as $\phi_{Ijk}$. Then $\{\phi_{Ijk} \}$ is an orthonormal basis in the Klein-Gordon norm of positive Rindler frequency solutions with support in ${\mathcal H}_I$. Furthermore, each $\phi_{Ijk}$ is composed of frequencies within an interval of size $\epsilon$ centered about the frequency $\omega_j \equiv (j + \frac{1}{2}) \epsilon$. Any real solution $g$ with support on ${\mathcal H}_I$ whose positive frequency part has finite Klein-Gordon norm can be expanded in this basis as
\begin{align}
    g = \sum_{j,k} g_{jk} \phi_{Ijk} + {\rm c.c}
\label{gdemp}
\end{align}
where ``c.c.'' denotes complex conjugation and $g_{jk} = \langle \phi_{Ijk} , g \rangle_{{\rm KG}}$. 

A corresponding orthonormal basis in the Klein-Gordon inner product for solutions with support on ${\mathcal H}_{II}$ can be obtained from $\{ \phi_{Ijk} \}$ as follows. Let $\mathcal R$ denote the reflection map on $\mathcal H$ about $\mathcal B$, i.e., the map defined by $V \to -V$. We define
\begin{align}
    \phi_{IIjk} = ({\mathcal R} \phi_{Ijk})^*
\end{align}
where the asterisk denotes complex conjugation. Then $\{ \phi_{IIjk} \}$ is an orthonormal basis of negative Rindler frequency solutions with support on ${\mathcal H}_{II}$. (As already noted, these solutions have positive Klein-Gordon norm on account of the reversal of Rindler time on ${\mathcal H}_{II}$.) The given solution $f$ can be expanded in this basis as 
\begin{align}
    f = \sum_{j,k} f_{jk} \phi_{IIjk} + {\rm c.c.}
\label{fdemp}
\end{align}
where $f_{jk} = \langle \phi_{IIjk} , f \rangle_{{\rm KG}}$.

Our task is to find $g$ so as to minimize the Klein-Gordon norm of the positive frequency part of $f+g$ with respect to affine parameter $V$. Equations~\eqref{gdemp}~and~\eqref{fdemp} provide decompositions of $g$ and $f$ into their positive and negative parts with respect to Rindler time. To obtain their positive and negative frequency parts with respect to affine time, we use the fact that the quantities
\begin{align}
\begin{aligned}
   F_{1jk} &= \frac{1}{\sqrt{1-e^{-2\pi \omega_j/\kappa}}}\left(\phi_{Ijk} + e^{-\pi \omega_j/\kappa} \phi^*_{IIjk} \right) \\
   F_{2jk} &= \frac{1}{\sqrt{1-e^{-2\pi \omega_j/\kappa}}}\left(\phi_{IIjk} + e^{-\pi \omega_j/\kappa} \phi^*_{Ijk} \right)
\end{aligned}
\label{F1F2}
\end{align}
are purely positive frequency\footnote{Here we ignore any errors resulting from the fact that $\phi_{Ijk}$ and $\phi^*_{IIjk}$ have a Rindler frequency spread of size $\epsilon$ about $\omega_j$, so that Eq.~\eqref{F1F2} will actually have a small negative frequency part. We will eventually take the limit as $\epsilon \to 0$.} with respect to affine time \cite{Wald_75,Wald_1995}. It follows immediately from the orthonormality and support properties of $\phi_{Ijk}$ and $\phi_{IIjk}$ that $\{ F_{1jk}, F_{2jk} \}$ are orthonormal in the Klein-Gordon norm. Furthermore, since
\begin{align}
\begin{aligned}
   \phi_{Ijk} &= \frac{1}{\sqrt{1-e^{-2\pi \omega_j/\kappa}}}\left(F_{1jk} - e^{-\pi \omega_j/\kappa} F^*_{2jk} \right) \\
   \phi_{IIjk} &= \frac{1}{\sqrt{1-e^{-2\pi \omega_j/\kappa}}}\left(F_{2jk} - e^{-\pi \omega_j/\kappa} F^*_{1jk} \right)
\end{aligned}
\end{align}
we may rewrite the basis expansions Eq.~\eqref{gdemp} and Eq.~\eqref{fdemp} as
\begin{align}
\begin{aligned}
   g &= \sum_{j,k} \frac{1}{\sqrt{1-e^{-2\pi \omega_j/\kappa}}} \left(g_{jk} F_{1jk} - g^*_{jk} e^{-\pi \omega_j/\kappa} F_{2jk} \right) + {\rm c.c.} \\
   f &= \sum_{j,k} \frac{1}{\sqrt{1-e^{-2\pi \omega_j/\kappa}}} \left(f_{jk} F_{2jk} - f^*_{jk} e^{-\pi \omega_j/\kappa} F_{1jk} \right) + {\rm c.c.}    
\end{aligned}
\label{fgdemp}
\end{align}
which provides a positive and negative frequency decomposition of $g$ and $f$ with respect to affine time.

Since $\{ F_{1jk}, F_{2jk} \}$ are orthonormal in the Klein-Gordon norm, it follows immediately that the Klein-Gordon norm of the positive frequency part of $f+g$ with respect to affine parameter is 
\begin{align}
||(f + g)^+||&^2_{{\rm KG}} = \sum_{j,k} \frac{1}{1-e^{-2\pi \omega_j/\kappa}} \times   \\ \nonumber 
&\times (|g_{jk} - e^{-\pi \omega_j/\kappa} f^*_{jk}|^2+ |f_{jk} - e^{-\pi \omega_j/\kappa} g^*_{jk}|^2)
\label{fgKG}
\end{align}

It is clear from this expression that for a given $f$, $||(f + g)^+||^2_{{\rm KG}}$ will be minimized if $g$ is chosen so that for each $j,k$, the quantity
\begin{align}
  &|g_{jk} - e^{-\pi \omega_j/\kappa}f^*_{jk}|^2 + |f_{jk} - e^{-\pi \omega_j/\kappa} g^*_{jk}|^2  \\
  =&~(|f_{jk}|^2+|g_{jk}|^2) (1+e^{-2\pi \omega_j/\kappa})-4e^{-\pi \omega_j/\kappa}\Re (f_{jk} g_{jk}) \nonumber 
\end{align} 
is minimized, where $\Re$ denotes the real part. The solution to this minimization problem for $g_{jk}$ is
\begin{align}
    g_{jk}  
     = \text{sech}(\pi \omega_j/\kappa) f^*_{jk}.
\label{gsol0}
\end{align}

This solution can be expressed in a more useful form as follows. Let $\tilde{f}$ be the reflection of $f$ about $V=0$. Thus, $\tilde{f}$ has support on ${\mathcal H}_I$ and is given by
\begin{align}
    \tilde{f}(V,y) = f(-V,y)
\end{align}
where $y$ denotes the transverse coordinates on $\mathcal H$. In terms of the boost Killing parameter $t$, we have
\begin{align}
    \tilde{f}(t,y) = f(t,y)
\end{align}
where on the left side $t$ denotes the Killing parameter on ${\mathcal H}_{I}$, whereas on the right side it denotes the Killing parameter on ${\mathcal H}_{II}$. We have
\begin{align}
   \tilde{f}_{jk} &\equiv \langle \phi_{Ijk} , \tilde{f} \rangle_{{\rm KG}} = - \langle {\mathcal R}\phi_{Ijk} , {\mathcal R} \tilde{f} \rangle_{{\rm KG}} \nonumber \\
   &= - \langle \phi^*_{IIjk}, f \rangle_{{\rm KG}} = \langle \phi_{IIjk}, f \rangle^*_{{\rm KG}} = f^*_{jk}.
\end{align}
Here, the minus sign in the second equality arises from the fact that $\mathcal R$ reverses the time direction. The sign change in the fourth equality arises from the explicit ``$i$'' in the Klein-Gordon product Eq.~\eqref{KGprod}. Thus, we have
\begin{align}
\label{eq:finalgf}
    g_{jk}  =\text{sech}(\pi \omega_j/\kappa)  \tilde{f}_{jk}.
\end{align}

Passing to the limit $\epsilon \to 0$ in our construction of $\phi_{jk}$ (so that $F_{1jk}$ and $F_{2jk}$ become exactly positive frequency with respect to affine time),\footnote{{More precisely, the relation Eq.~\eqref{ghatsol} is obtained by, for each $\omega$, letting $\epsilon = \omega/n$ and $j=n$ where $n$ is an integer and considering the limit as $n\to \infty$.  }} we see that the optimal $g$ is obtained as follows: Given $f$ with support on ${\mathcal H}_{II}$, reflect it about the bifurcation surface $V=0$ to get a corresponding function $\tilde{f}$ on ${\mathcal H}_I$. Fourier transform $\tilde{f}$ with respect to Rindler time $t$ and define $g$ via its Fourier transform by
\begin{align}
    \hat{g}(\omega, y)  = \text{sech}(\pi \omega/\kappa) \widehat{\tilde{f}}(\omega, y) .
\label{ghatsol}
\end{align}
Note that in the high frequency limit $\omega \gg \kappa$, we have $\hat{g}(\omega, y) \to 0$, whereas in the low frequency limit $\omega \ll \kappa$, we have $\hat{g}(\omega, y) \to \widehat{\tilde{f}}(\omega, y)$. Thus, one may view the optimal purification as corresponding to sending in time-reversed radiation modified by a cutoff for Rindler frequencies of order $\kappa$ and above.

The inverse Fourier transform of the factor appearing in Eq.~\eqref{ghatsol} is 
\begin{align}
    \int_{-\infty}^{\infty} \frac{d\omega}{2\pi}~\text{sech}(\pi \omega/\kappa) e^{{-}i\omega t} = \frac{\kappa}{2\pi}~ \text{sech}\left[{\kappa t}/2\right].
    \label{eq:FTfactor}
\end{align}
Since a product in Fourier transform space corresponds to a convolution in position space, we have 
\begin{align}
    g(t,y) = \frac{\kappa}{2\pi} \int_{-\infty}^\infty \text{sech}\left[{ \kappa} (t-t')/2 \right] \tilde{f}(t',y) dt'.
\label{gsol}
\end{align}

Let us now consider how much improvement is obtained by using this optimal recovery protocol rather than doing nothing. It can be seen from Eq.~\eqref{fgKG} that, for $g=0$, we have
\begin{align}\begin{aligned}
\label{eq:fnorm}
       ||f^+||^2_{{\rm KG}} &= \frac{1}{\pi}\int\limits_{\mathbb{R}^{d-2}}d^{d-2}y\int_{0}^{\infty} d\omega ~\omega  \coth(\pi \omega/\kappa)|\hat{f}(\omega,y) |^2 .
     \end{aligned}
\end{align}
On the other hand, for the optimal $g$ given by Eq.~\eqref{ghatsol}, we have 
\begin{align}\begin{aligned}
      ||(f &+g)^+||^2_{{\rm KG}}  \\&= \frac{1}{\pi}\int\limits_{\mathbb{R}^{d-2}}d^{d-2}y\int_{0}^{\infty} d\omega ~\omega \tanh(\pi \omega/\kappa) |\hat{f}(\omega,y) |^2
       \end{aligned}
\end{align}
Thus, the improvement obtained by using the optimal $g$ is negligible at high frequencies, $\omega \gg \kappa$, but is very significant at low frequencies, $\omega \ll \kappa$. {The corresponding optimal purification of the Maxwell and the linearized gravitational field is given by a similar formula to Eq.~\eqref{eq:finalgf} in terms of the ``free data'' on $\mc{H}$ (see, e.g., \cite{2022IJMPD..3141003D,Danielson2022sga} for further details). }

Finally, we have assumed above that $f$ has support in region II of Fig. \ref{fig:bifurcate_tikz} and, in particular, vanishes at the bifurcation surface, $\mathcal B$. This assumption was needed for the positive Rindler frequency part of $f$ to have finite Klein-Gordon norm. Indeed, if $f$ did not go to zero as $t \to -\infty$ (i.e., at $\mathcal B$), then, assuming that $f \to 0$ as $t \to \infty$ (i.e., $V \to - \infty$), there would be infrared divergences in the Klein-Gordon norm of the positive frequency part of $f$. A forthcoming article by one of us \cite{IDT} provides a general treatment of these difficulties, but in the present analysis the finite Klein-Gordon norm of $f$ was necessary to get the well-defined basis expansion Eq.~\eqref{fdemp} of $f$ in the modes $\phi_{Ijk}$. This, in turn, was used to get a well defined basis expansion of $f$ in the positive affine frequency modes $\{ F_{1jk}, F_{2jk} \}$. 

Nevertheless, it is of interest to find the optimal continuation of $f$ into region I in the case where $f$ does not vanish at $\mathcal B$. The solution Eq.~\eqref{gsol} for the continuation of $f$ into ${\mathcal H}_I$ that we obtained in the case where $f$ vanishes at $\mathcal B$ continues to be well defined in the case where $f$ does not vanish at $\mathcal B$. We believe that this could be proven to be the optimal continuation in the general case by the following argument. Given $f(V, y)$ for $V\leq 0$, which does not necessarily vanish at $V=0$, we wish to determine the extension $g(V,y)$ for $V > 0$ so as to minimize the Klein-Gordon norm of the positive affine frequency part of $f+g$. We consider the inner product of $f+g$ with the orthonormal mode functions $\{ F_{1jk}, F_{2jk} \}$. The minimization problem for the magnitude of each of these inner products is identical to the problem we have just solved. Thus, the optimal solution for $V > 0$ is given by Eq.~\eqref{gsol0}. If $\{ F_{1jk}, F_{2jk} \}$ were a basis for the Hilbert space of the all positive affine frequency solutions\footnote{This Hilbert space is obtained by starting with smooth functions of compact support on $\mathcal H$ and taking their completion in the Klein-Gordon norm of their positive frequency parts.} on $\mathcal H$, then we have solved the optimization problem and the solution would be given by Eq.~\eqref{gsol}. However, if $\{ F_{1jk}, F_{2jk} \}$ is not a basis, then we have only solved the optimization problem on a proper subset of possible continuations of $f$. Thus, the key issue is whether $\{ F_{1jk}, F_{2jk} \}$ is a basis. Equation Equation~\eqref{fgdemp} essentially shows that it is a basis for the positive affine frequency part arising from any smooth function of compact support on $\mathcal H$ that vanishes at $V$. Thus, the key issue reduces to the question of whether the subspace of positive affine frequency solutions that arise from taking the positive affine frequency part of smooth functions of compact support that vanish at $V=0$ are dense (in the Klein-Gordon norm) in the space of positive affine frequency solutions that arise from taking the positive affine frequency part of smooth functions of compact support with no restriction at $V=0$. Lemma 5.1 of \cite{Brasco_2020} indicates that this is the case. The norm considered in that lemma for the case $s= 1/2$ and $n=1$ corresponds to the contribution to the Klein-Gordon norm from a single generator of $\mathcal H$. The lemma shows that one can find smooth, real-valued functions of compact support on $\mathbb{R}$ that are equal to $1$ at $V=0$ but have arbitrarily small Klein-Gordon norm of their positive frequency part. It follows that any smooth function of compact support on $\mathcal H$ can be written as a sum of a smooth function of compact support that vanishes on $\mathcal B$ plus a smooth function of compact support whose positive affine frequency part has an arbitrarily small Klein-Gordon norm. This implies that $\{ F_{1jk}, F_{2jk} \}$ is a basis of the entire Klein-Gordon Hilbert space.

In the remainder of this paper, we will assume that, as we have just argued, Eq.~\eqref{gsol} remains the optimal solution in the case where $f$ does not necessarily vanish at $V=0$.

\section{minimizing the decoherence outside a black hole}
\label{bhmin}

We now return to the gedankenexperiment of \cite{2022IJMPD..3141003D,Danielson2022sga}, as discussed in the introduction. Alice is in the midst of performing her experiment when, at $t=t_c$, she receives a phone call from central headquarters telling her to abort her experiment and preserve as much coherence of her particle as possible. Since some entangling radiation has already entered the black hole as a consequence of her earlier actions, it will not be possible for Alice to perfectly restore the coherence of her particle. What is the maximal coherence Alice can, in principle, achieve? And what should Alice do for $t > t_c$ to try to achieve this maximal coherence?

As we previously argued in \cite{2022IJMPD..3141003D,dswLocal_2024}, Alice can minimize her decoherence due to radiating to infinity by performing her separation and recombination sufficiently slowly. We will therefore neglect any radiation to infinity and will focus entirely on the radiation falling into the black hole. Let $\mathcal C$ denote the cross-section of the event horizon $\mathcal H$ corresponding to the intersection of the horizon with the boundary of the future of the event in her laboratory when she received the phone call, as illustrated in Fig.~\ref{fig:BH_tikz}. Without loss of generality, we may choose the Killing parameter, $t$, on $\mathcal H$ so that $\mathcal C$ corresponds to the time $t = t_c$ on $\mathcal H$. The corresponding affine time of $\mathcal C$ is then $V = V_c$ where
\begin{align}
    V_c = e^{\kappa t_c}.
\end{align}
Let $f$ be the restriction to the event horizon, $\mathcal H$, of the classical solution corresponding to the difference between the retarded solutions corresponding to the two paths of Alice's particle. The coherent state associated with $f$ then describes the entangling radiation that enters the black hole. When Alice receives her phone call, she can do nothing about $f$ for $t < t_c$. However, she can, to a large degree, control the entangling radiation that enters the black hole for $t > t_c$. How should she continue her experiment so as to minimize the decoherence? Specifically, if we imagine that she can perfectly control the entangling radiation emitted by her particle, given $f$ on $\mathcal H$ for $t \leq t_c$, what continuation, $g$, of this radiation for $t > t_c$ should she choose so as to minimize the decoherence? 

This question is very similar to the mathematical question posed and answered in the previous section. The only significant difference is that the bifurcation surface, $\mathcal B$, in Sec.~\ref{rindler} has now been replaced by a cross-section, $\mathcal C$, that lies to the future of $\mathcal B$. It might seem that this is a very significant difference because the decomposition of $f$ and $g$ into positive and negative Killing frequencies played a crucial role in the analysis. Such a decomposition made sense in our Rindler problem because $f$ and $g$ were both defined over the entire range $-\infty < t < \infty$. However, such a decomposition would not make sense here since $f$ is defined only for $t \leq t_c$ and $g$ is defined only for $t > t_c$. 

Nevertheless, the fact that $t$ is the Killing parameter of an exact Killing field on spacetime was not actually needed for the analysis of the optimal $g$ given in the previous section. All that was actually used is that the horizon geometry itself is invariant under affine translations and the Killing parameter $t$ is related to affine parameter $V$ via Eqs.~\eqref{ka1} and~\eqref{ka2}. Therefore, we can proceed in the present case by defining a ``pseudo-Killing parameter'' $v$ on the black hole event horizon by\footnote{Note that exactly the same analysis could be applied to determine how to minimize decoherence due to radiation to null infinity, given radiation $f$ up to a cross-section $\mathcal C$ at $V=V_c$, where $V$ now denotes affine time at null infinity. At null infinity, Killing time $t$ and affine time $V$ coincide, but one could nevertheless define a ``pseudo Killing parameter'' $v$ at null infinity by Eq.~(\ref{eq:pseudoKill}) (where $\kappa$ now is chosen arbitrarily). The optimal choice of $g$ would then be given by Eq. (\ref{gsol2}), which could be reexpressed in terms of affine time (i.e., Killing time) by Eq. (\ref{gsol3}), where the limits of the integral now extend from $V_c$ to $\infty$.}
\begin{align}
\begin{aligned}
\label{eq:pseudoKill}
   V - V_c &= e^{\kappa v} \quad \quad \quad V > V_c \\
   V - V_c &= - e^{\kappa v} \quad \quad \, \, V < V_c
\end{aligned}.
\end{align}
We view $f$ for $V < V_c$ as a function of $v$ and decompose it into its positive and negative frequency parts with respect to $v$. This choice of positive frequency on the future horizon corresponds to Alice performing her experiment in the Hartle-Hawking vacuum. As we have previously argued in \cite{2022IJMPD..3141003D,dswLocal_2024}, at low Killing frequencies $\omega\ll \kappa$, the Unruh state is equivalent to the Hartle-Hawking state on $\mc{H}^{+}$, and so our results will also yield the optimal purification in that vacuum if Alice performs her experiment sufficiently adiabatically. The optimal choice of $g$ for $V > V_c$ to minimize the Klein-Gordon norm of the positive affine frequency part of $f+g$ then follows immediately from the results of the previous section. Specifically, by Eq.~\eqref{gsol}, the optimal choice of $g$ is given by
\begin{align}
    g(v,y) = \frac{\kappa}{2\pi} \int_{-\infty}^\infty \text{sech}\left[{\kappa}(v-v')/2 \right] \tilde{f}(v',y) dv'
\label{gsol2}
\end{align}
where $\tilde{f}$ is the reflection of $f$ about $V = V_c$, and $y$ now denotes the angular coordinates on $\mathcal H$. Although we have taken the upper limit of this integral to be $+\infty$, $f$ can be nonzero only for $V > 0$ (since the entire range $-\infty < t < \infty$ of true Killing time $t$ in Alice's laboratory occurs for $V > 0$), so the upper limit may be replaced by $\kappa^{-1} \ln V_c$. 
\begin{figure}
    \centering
\includegraphics[width=\linewidth]{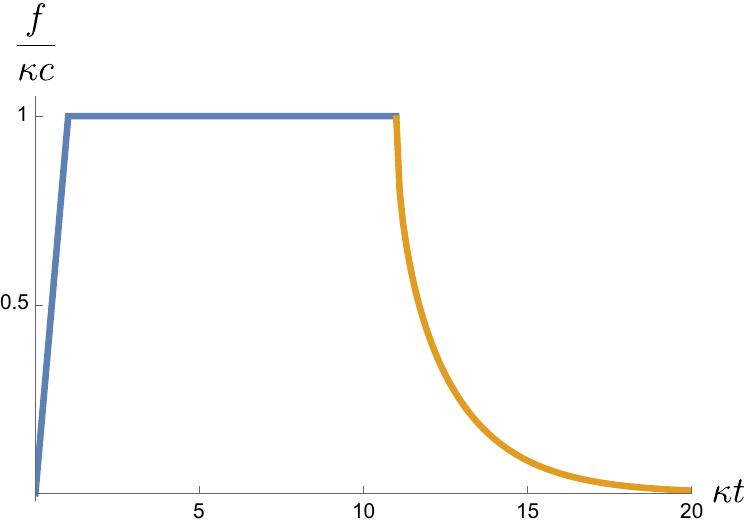}
    \caption{The optimal ``ramp down'' is shown in orange for the example of Eq.~\eqref{traj1} for the parameters $t_1 = \kappa^{-1}$ and $t_c = 11\kappa^{-1}$.}
    \label{fig:recoveries2}
\end{figure}
This result can be rewritten in terms of affine time $V>V_c$ as
\begin{align}
    g(V,y) 
     &=\frac{1}{\pi}\int_{V_c}^{2V_c} {}\frac{ f(2V_c - V',y)}{V-2 V_c+V'}\sqrt{\frac{V-V_c}{V'-V_c}}dV'
\label{gsol3}
\end{align}
where in this formula, we view $f$ and $g$ as functions of $V$ (rather than $v$), and we have used the fact that for $V > V_c$, we have $\tilde{f}(V, y) = f(2V_c - V, y)$.
This can, in turn, be written in terms of the true Killing time $t$ for $t > t_c$ by substituting $V = e^{\kappa t}$. Viewing $f$ and $g$ now as functions of $t$, we have
\begin{align}
\label{gsol4}
    g(t,y) 
   =
   \frac{\kappa}{\pi}\int_{t_c}^{t_c + \kappa^{-1} \ln 2} &\bigg[\frac{e^{\kappa  t'} f(\kappa^{-1} \ln[2e^{\kappa t_c} - e^{\kappa t'}],y) }{e^{\kappa  t}-2 e^{\kappa  t_c}+e^{\kappa 
   t'}}\times 
   \\ \nonumber 
   &\times \sqrt{\frac{e^{\kappa  t_c}-e^{\kappa  t}}{e^{\kappa 
   t_c}-e^{\kappa  t'}}}\bigg]dt'.
\end{align}

It should be noted that even if Alice has completed her experiment at the time $t=t_c$---i.e., even if $f(t_c) = 0$---the optimal strategy for Alice to minimize decoherence is for her to reopen the superposition and perform further manipulations so as to produce the radiation given by Eq.~\eqref{gsol4}. Nevertheless, it also can be seen from Eq.~\eqref{gsol4} that for $t - t_c \gg \kappa^{-1}$, we have
\begin{figure}
    \centering
    \includegraphics[width=\linewidth]{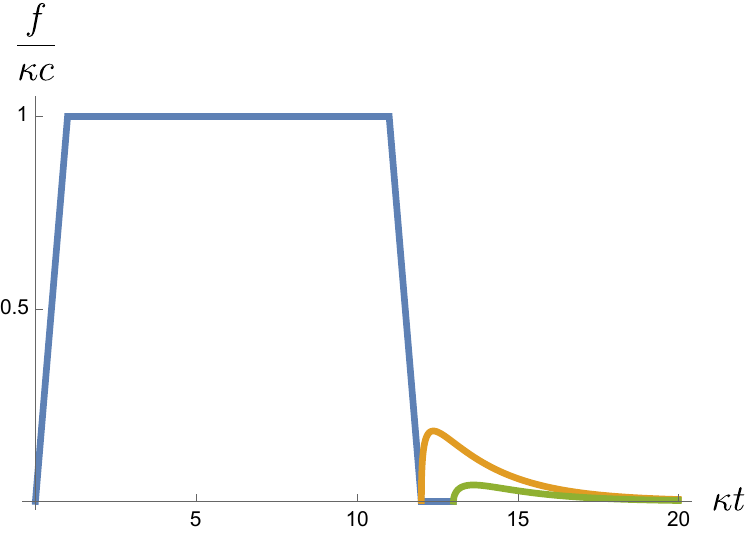}
    \caption{ The optimal radiation, $g$, for the choice of $f$ given by Eq.~\eqref{traj2}, with parameters $t_1 = \kappa^{-1}$ and $T= 10\kappa^{-1}$. The orange curve is for $t_c = 2t_1 +T$ (``immediate recovery''), and the green curve is for $t_c = 2t_1 +T + \kappa^{-1}$ (recovery after a delay time of $\kappa^{-1}$).}
    \label{fig:recoveries}
\end{figure}
\begin{align}
    |g(t,y)| \leq \frac{1}{\pi}e^{-\kappa(t - t_c)} \, {\rm max} |f| .
\label{gdec}
\end{align}
Thus, the optimal radiation is very small except for a time interval of size $\sim \kappa^{-1}$ after time $t_c$. In particular, if Alice holds her superposition open for a time $ T \gg \kappa^{-1}$ as contemplated in the gedankenexperiments of \cite{2022IJMPD..3141003D,Danielson2022sga}, it is clear that the improvement provided by Eq.~\eqref{gsol4} is much too small to affect the estimates of decoherence given in \cite{2022IJMPD..3141003D,Danielson2022sga}.

Our optimal solution Eq.~\eqref{gsol4} and the improvement it gives is best illustrated by giving several explicit examples. For the first example, suppose that Alice opens her superposition between $t=0$ and $t=t_1$ and then holds the superposition stationary until time $t_c$, when she is instructed to minimize the decoherence. For simplicity and definiteness, we will suppose that $f$ takes the following simple form:
\begin{align} 
    f(t,y) = \begin{cases}
        0, & t < 0,
        \\
        ct, & 0 < t < t_1
        \\
        ct_1, & t_1 < t \leq t_c
    \end{cases}.
\label{traj1}
\end{align}
 The optimal choice of ``ramp down'' for $t > t_c$ is shown in Fig.~\ref{fig:recoveries2} for a particular choice of the parameters. {As illustrated in the figure, the ramp down actually begins with an infinite negative slope at $t=t_c$ and then transitions to an} exponential decay with time scale set by $\kappa^{-1}$ [see Eq.~\eqref{gdec}].

 As a second example, suppose Alice opens her superposition as in the previous example, but she now leaves her superposition open only for a time $T$ and then closes the superposition over a time interval $t_1$. For simplicity and definiteness, we will suppose that $f$ takes the following simple form:
\begin{align}
    f(t,y) = \begin{cases}
        0 , & t< 0,
        \\
        ct, & t_0<t<t_1
        \\
        ct_1, & t_1 < t< t_1 + T
        \\
        c(2t_1+T-t), & t_1 + T < t < 2t_1 + T 
        \\
        0, & 2t_1 + T < t \leq t_c
    \end{cases}.
\label{traj2}
\end{align}

The optimal choice of $g$ is shown in Fig.~\ref{fig:recoveries} for a particular choice of the parameters of $f$ and two different choices of $t_c$. The orange curve shows the optimal $g$ in the case where {$t_c = 2t_1 + T$; i.e.,} $t_c$ is taken to be the moment that the original superposition was closed. The green curve shows the optimal $g$ for the case where {$t_c = 2t_1 + T + \kappa^{-1}$; i.e.,} $t_c$ is a time $\kappa^{-1}$ after the original superposition was closed. The key point to note is that, although in both cases the superposition has already been closed by time $t_c$, Alice can improve the coherence of her particle by reopening the superposition and emitting some additional entangling radiation into the black hole. In the first case (orange curve), where she acts immediately {after the superposition was closed, her further actions will decrease $\langle N \rangle$ by $1.3 \%$. In the second case (green curve)} where she is unable to act until a time $\kappa^{-1}$ after the original superposition was closed, then {her further actions will decrease $\langle N \rangle$ by only $0.07\%$. It is worth noting that if Alice were able to act optimally beginning at time $t_c = t_1 + T$ (i.e., just before the start of the linear ramp down of $f$), then a decrease of $\langle N \rangle$ of $4.6 \%$ would occur as compared with what occurs with the linear ramp down in Eq.~\eqref{traj2}. All of these percentage improvements in $\langle N \rangle$ decrease with $T$ and go to zero as $T \to \infty$.}

\section{Discussion}
\label{disc}

The optimal purification that we have found has an interesting form. At low frequencies ($\omega \ll 1/\kappa$), it corresponds to performing a reflection about the bifurcation surface $\mathcal B$ (in the Rindler case) or the cross-section $\mathcal C$ (in the black hole case), which we refer to as a ``CRT purification.'' At high frequencies ($\omega \gg 1/\kappa$), it corresponds to leaving the vacuum state unperturbed, which we refer to as a ``Minkowski purification.'' At intermediate frequencies, it smoothly interpolates between these. The CRT purification is familiar from algebraic quantum field theory. When performing the GNS construction of the Hilbert space over an algebraic state, the CRT purification is the state that corresponds to the identity operator. In the language of Tomita-Takesaki theory, this is the vector representative of the coherent state in the so-called ``natural cone'' of the vacuum state. The absolute value of the inner product between this state and the global vacuum is a well-behaved ``fidelity'' first studied by Holevo \cite{kholevo1972quasiequivalence}. In terms of modes, with the CRT purification, the total number of particles in the states is
\begin{align}\begin{aligned}
       \braket{N}_{\rm CRT} &=\frac{2}{\pi}\int\limits_{\mathbb{R}^{d-2}}d^{d-2}y\int_{0}^{\infty} d\omega~\omega  \tanh\left(\frac{\pi \omega}{2\kappa}\right) |f_{\omega }(y) |^2.
       \end{aligned}
\end{align}
This regulates the IR divergence at small $\omega$, just as the optimal purification. However, at high energies, there are twice the number of particles in the state as in the optimal purification (or nonrecovered) state. This is because at high energies, the CRT reflection is simply putting an extra copy of particles in a different position. 
The Holevo fidelity may be contrasted with the more commonly studied Uhlmann fidelity \cite{uhlmann1976transition}, which corresponds to the overlap of the optimal purification with the vacuum. Obviously, the Uhlmann fidelity is lower bounded by Holevo's fidelity, and our calculations suggest that they converge for low energy processes. 

Finally, we note that there is a fundamental connection between the fidelity of Alice's experiment and the extent to which an observer, living behind the horizon, can gain information about Alice's experiment. This notion of complementarity has been formalized in the quantum information theory literature, going under the name of the information-disturbance tradeoff (IDT). {There are several formulations of the IDT, the most relevant to the present work being formulated in terms of fidelities \cite{2010PhRvL.104l0501B}
\begin{align}
    \max_{\mathcal{R}} F(\mathcal{R} \circ \mathcal{N}, \text{Id}) = \max_{\mathcal{R}'} F( \mathcal{N}_c, \mathcal{R'}).
\end{align}
Here, the fidelities are between quantum channels, which are completely positive, trace-preserving maps on quantum states. The quantum channel fidelity is equal to the Uhlmann fidelity of states outputted by the channels, minimized over input states. In the present setting, $\mathcal{N}$ corresponds to the channel sending the state of Alice's particle prior to the experiment to the state of her particle after recombination. The complementary channel, $\mathcal{N}_c$, is the map from Alice's particle's initial state to the final state of the radiation on the black hole {{horizon}}. The maximization is over recovery channels, $\mathcal{R}$, of which we found the 
optimal channel in the space of coherent states characterized by the function $g$. The maximization of $\mathcal{R}'$ is over channels that take all states to some fixed state, which in our context is the vacuum on the horizon. The IDT quantifies the relation between decoherence of Alice's particle and distinguishability of the quantum states of the black hole. Alice's experiment {{decoheres}} if and only if it sends information to the black hole. This connection {can be used to teleport information into the black hole interior using soft quanta \cite{Kudler-Flam:2025yur} and} will be explored in detail in a forthcoming paper by one of us \cite{IDT}. }

\begin{acknowledgments}
We wish to thank Peter Hintz for bringing reference \cite{Brasco_2020} to our attention and for helpful discussions concerning that result.
J.K.F. is supported by the Marvin L.~Goldberger Member Fund at the Institute for Advanced Study and the National Science Foundation under Grant No. PHY-2207584. D.L.D. acknowledges support as a Fannie and John Hertz Foundation Fellow holding the Barbara Ann Canavan Fellowship and as an Eckhardt Graduate Scholar at the University of Chicago. D.L.D. and R.M.W. were supported, in part, by NSF Grant No. PHY-2403584 and Templeton Foundation Grant No. 62845 to the University of Chicago. G.S. is supported by the Princeton Gravity Initiative at Princeton University.
\end{acknowledgments}

\section{Data Availability}
No data were created or analyzed in this study.

\normalem
\bibliography{main}

\begin{thebibliography}{40}%
\makeatletter
\providecommand \@ifxundefined [1]{%
 \@ifx{#1\undefined}
}%
\providecommand \@ifnum [1]{%
 \ifnum #1\expandafter \@firstoftwo
 \else \expandafter \@secondoftwo
 \fi
}%
\providecommand \@ifx [1]{%
 \ifx #1\expandafter \@firstoftwo
 \else \expandafter \@secondoftwo
 \fi
}%
\providecommand \natexlab [1]{#1}%
\providecommand \enquote  [1]{``#1''}%
\providecommand \bibnamefont  [1]{#1}%
\providecommand \bibfnamefont [1]{#1}%
\providecommand \citenamefont [1]{#1}%
\providecommand \href@noop [0]{\@secondoftwo}%
\providecommand \href [0]{\begingroup \@sanitize@url \@href}%
\providecommand \@href[1]{\@@startlink{#1}\@@href}%
\providecommand \@@href[1]{\endgroup#1\@@endlink}%
\providecommand \@sanitize@url [0]{\catcode `\\12\catcode `\$12\catcode
  `\&12\catcode `\#12\catcode `\^12\catcode `\_12\catcode `\%12\relax}%
\providecommand \@@startlink[1]{}%
\providecommand \@@endlink[0]{}%
\providecommand \url  [0]{\begingroup\@sanitize@url \@url }%
\providecommand \@url [1]{\endgroup\@href {#1}{\urlprefix }}%
\providecommand \urlprefix  [0]{URL }%
\providecommand \Eprint [0]{\href }%
\providecommand \doibase [0]{https://doi.org/}%
\providecommand \selectlanguage [0]{\@gobble}%
\providecommand \bibinfo  [0]{\@secondoftwo}%
\providecommand \bibfield  [0]{\@secondoftwo}%
\providecommand \translation [1]{[#1]}%
\providecommand \BibitemOpen [0]{}%
\providecommand \bibitemStop [0]{}%
\providecommand \bibitemNoStop [0]{.\EOS\space}%
\providecommand \EOS [0]{\spacefactor3000\relax}%
\providecommand \BibitemShut  [1]{\csname bibitem#1\endcsname}%
\let\auto@bib@innerbib\@empty
\bibitem [{\citenamefont {Bose}\ \emph {et~al.}(2017)\citenamefont {Bose},
  \citenamefont {Mazumdar}, \citenamefont {Morley}, \citenamefont {Ulbricht},
  \citenamefont {Toro\v{s}}, \citenamefont {Paternostro}, \citenamefont
  {Geraci}, \citenamefont {Barker}, \citenamefont {Kim},\ and\ \citenamefont
  {Milburn}}]{Bose_2017}%
  \BibitemOpen
  \bibfield  {author} {\bibinfo {author} {\bibfnamefont {S.}~\bibnamefont
  {Bose}}, \bibinfo {author} {\bibfnamefont {A.}~\bibnamefont {Mazumdar}},
  \bibinfo {author} {\bibfnamefont {G.~W.}\ \bibnamefont {Morley}}, \bibinfo
  {author} {\bibfnamefont {H.}~\bibnamefont {Ulbricht}}, \bibinfo {author}
  {\bibfnamefont {M.}~\bibnamefont {Toro\v{s}}}, \bibinfo {author}
  {\bibfnamefont {M.}~\bibnamefont {Paternostro}}, \bibinfo {author}
  {\bibfnamefont {A.~A.}\ \bibnamefont {Geraci}}, \bibinfo {author}
  {\bibfnamefont {P.}~\bibnamefont {Barker}}, \bibinfo {author} {\bibfnamefont
  {M.~S.}\ \bibnamefont {Kim}},\ and\ \bibinfo {author} {\bibfnamefont
  {G.}~\bibnamefont {Milburn}},\ }\bibfield  {title} {\bibinfo {title} {{Spin
  Entanglement Witness for Quantum Gravity}},\ }\href
  {https://doi.org/10.1103/PhysRevLett.119.240401} {\bibfield  {journal}
  {\bibinfo  {journal} {Phys. Rev. Lett.}\ }\textbf {\bibinfo {volume} {119}},\
  \bibinfo {pages} {240401} (\bibinfo {year} {2017})},\ \Eprint
  {https://arxiv.org/abs/1707.06050} {arXiv:1707.06050 [quant-ph]} \BibitemShut
  {NoStop}%
\bibitem [{\citenamefont {Marletto}\ and\ \citenamefont
  {Vedral}(2017)}]{Marletto_2017}%
  \BibitemOpen
  \bibfield  {author} {\bibinfo {author} {\bibfnamefont {C.}~\bibnamefont
  {Marletto}}\ and\ \bibinfo {author} {\bibfnamefont {V.}~\bibnamefont
  {Vedral}},\ }\bibfield  {title} {\bibinfo {title} {{Gravitationally-induced
  entanglement between two massive particles is sufficient evidence of quantum
  effects in gravity}},\ }\href
  {https://doi.org/10.1103/PhysRevLett.119.240402} {\bibfield  {journal}
  {\bibinfo  {journal} {Phys. Rev. Lett.}\ }\textbf {\bibinfo {volume} {119}},\
  \bibinfo {pages} {240402} (\bibinfo {year} {2017})},\ \Eprint
  {https://arxiv.org/abs/1707.06036} {arXiv:1707.06036 [quant-ph]} \BibitemShut
  {NoStop}%
\bibitem [{\citenamefont {Belenchia}\ \emph {et~al.}(2018)\citenamefont
  {Belenchia}, \citenamefont {Wald}, \citenamefont {Giacomini}, \citenamefont
  {Castro-Ruiz}, \citenamefont {Brukner},\ and\ \citenamefont
  {Aspelmeyer}}]{Belenchia_2018}%
  \BibitemOpen
  \bibfield  {author} {\bibinfo {author} {\bibfnamefont {A.}~\bibnamefont
  {Belenchia}}, \bibinfo {author} {\bibfnamefont {R.~M.}\ \bibnamefont {Wald}},
  \bibinfo {author} {\bibfnamefont {F.}~\bibnamefont {Giacomini}}, \bibinfo
  {author} {\bibfnamefont {E.}~\bibnamefont {Castro-Ruiz}}, \bibinfo {author}
  {\bibfnamefont {v.}~\bibnamefont {Brukner}},\ and\ \bibinfo {author}
  {\bibfnamefont {M.}~\bibnamefont {Aspelmeyer}},\ }\bibfield  {title}
  {\bibinfo {title} {{Quantum Superposition of Massive Objects and the
  Quantization of Gravity}},\ }\href
  {https://doi.org/10.1103/PhysRevD.98.126009} {\bibfield  {journal} {\bibinfo
  {journal} {Phys. Rev. D}\ }\textbf {\bibinfo {volume} {98}},\ \bibinfo
  {pages} {126009} (\bibinfo {year} {2018})},\ \Eprint
  {https://arxiv.org/abs/1807.07015} {arXiv:1807.07015 [quant-ph]} \BibitemShut
  {NoStop}%
\bibitem [{\citenamefont {Christodoulou}\ and\ \citenamefont
  {Rovelli}(2019)}]{Christodoulou_2019}%
  \BibitemOpen
  \bibfield  {author} {\bibinfo {author} {\bibfnamefont {M.}~\bibnamefont
  {Christodoulou}}\ and\ \bibinfo {author} {\bibfnamefont {C.}~\bibnamefont
  {Rovelli}},\ }\bibfield  {title} {\bibinfo {title} {On the possibility of
  laboratory evidence for quantum superposition of geometries},\ }\href
  {https://doi.org/https://doi.org/10.1016/j.physletb.2019.03.015} {\bibfield
  {journal} {\bibinfo  {journal} {Physics Letters B}\ }\textbf {\bibinfo
  {volume} {792}},\ \bibinfo {pages} {64} (\bibinfo {year} {2019})}\BibitemShut
  {NoStop}%
\bibitem [{\citenamefont {Giacomini}\ \emph {et~al.}(2019)\citenamefont
  {Giacomini}, \citenamefont {Castro-Ruiz},\ and\ \citenamefont
  {Brukner}}]{Giacomini_2019}%
  \BibitemOpen
  \bibfield  {author} {\bibinfo {author} {\bibfnamefont {F.}~\bibnamefont
  {Giacomini}}, \bibinfo {author} {\bibfnamefont {E.}~\bibnamefont
  {Castro-Ruiz}},\ and\ \bibinfo {author} {\bibfnamefont {v.}~\bibnamefont
  {Brukner}},\ }\bibfield  {title} {\bibinfo {title} {{Quantum mechanics and
  the covariance of physical laws in quantum reference frames}},\ }\href
  {https://doi.org/10.1038/s41467-018-08155-0} {\bibfield  {journal} {\bibinfo
  {journal} {Nature Commun.}\ }\textbf {\bibinfo {volume} {10}},\ \bibinfo
  {pages} {494} (\bibinfo {year} {2019})},\ \Eprint
  {https://arxiv.org/abs/1712.07207} {arXiv:1712.07207 [quant-ph]} \BibitemShut
  {NoStop}%
\bibitem [{\citenamefont {Gonzalez-Ballestero}\ \emph
  {et~al.}(2021)\citenamefont {Gonzalez-Ballestero}, \citenamefont
  {Aspelmeyer}, \citenamefont {Novotny}, \citenamefont {Quidant},\ and\
  \citenamefont {Romero-Isart}}]{Aspelmeyer_2021}%
  \BibitemOpen
  \bibfield  {author} {\bibinfo {author} {\bibfnamefont {C.}~\bibnamefont
  {Gonzalez-Ballestero}}, \bibinfo {author} {\bibfnamefont {M.}~\bibnamefont
  {Aspelmeyer}}, \bibinfo {author} {\bibfnamefont {L.}~\bibnamefont {Novotny}},
  \bibinfo {author} {\bibfnamefont {R.}~\bibnamefont {Quidant}},\ and\ \bibinfo
  {author} {\bibfnamefont {O.}~\bibnamefont {Romero-Isart}},\ }\bibfield
  {title} {\bibinfo {title} {Levitodynamics: Levitation and control of
  microscopic objects in vacuum},\ }\href
  {https://doi.org/10.1126/science.abg3027} {\bibfield  {journal} {\bibinfo
  {journal} {Science}\ }\textbf {\bibinfo {volume} {374}},\ \bibinfo {pages}
  {3027} (\bibinfo {year} {2021})},\ \Eprint {https://arxiv.org/abs/2111.05215}
  {arXiv:2111.05215 [quant-ph]} \BibitemShut {NoStop}%
\bibitem [{\citenamefont {Danielson}\ \emph {et~al.}(2022)\citenamefont
  {Danielson}, \citenamefont {Satishchandran},\ and\ \citenamefont
  {Wald}}]{Danielson:2021egj}%
  \BibitemOpen
  \bibfield  {author} {\bibinfo {author} {\bibfnamefont {D.~L.}\ \bibnamefont
  {Danielson}}, \bibinfo {author} {\bibfnamefont {G.}~\bibnamefont
  {Satishchandran}},\ and\ \bibinfo {author} {\bibfnamefont {R.~M.}\
  \bibnamefont {Wald}},\ }\bibfield  {title} {\bibinfo {title}
  {{Gravitationally mediated entanglement: Newtonian field versus gravitons}},\
  }\href {https://doi.org/10.1103/PhysRevD.105.086001} {\bibfield  {journal}
  {\bibinfo  {journal} {Phys. Rev. D}\ }\textbf {\bibinfo {volume} {105}},\
  \bibinfo {pages} {086001} (\bibinfo {year} {2022})},\ \Eprint
  {https://arxiv.org/abs/2112.10798} {arXiv:2112.10798 [quant-ph]} \BibitemShut
  {NoStop}%
\bibitem [{\citenamefont {Carney}(2022)}]{Carney_2021}%
  \BibitemOpen
  \bibfield  {author} {\bibinfo {author} {\bibfnamefont {D.}~\bibnamefont
  {Carney}},\ }\bibfield  {title} {\bibinfo {title} {{Newton, entanglement, and
  the graviton}},\ }\href {https://doi.org/10.1103/PhysRevD.105.024029}
  {\bibfield  {journal} {\bibinfo  {journal} {Phys. Rev. D}\ }\textbf {\bibinfo
  {volume} {105}},\ \bibinfo {pages} {024029} (\bibinfo {year} {2022})},\
  \Eprint {https://arxiv.org/abs/2108.06320} {arXiv:2108.06320 [quant-ph]}
  \BibitemShut {NoStop}%
\bibitem [{\citenamefont {Christodoulou}\ \emph {et~al.}(2023)\citenamefont
  {Christodoulou}, \citenamefont {Di~Biagio}, \citenamefont {Aspelmeyer},
  \citenamefont {Brukner}, \citenamefont {Rovelli},\ and\ \citenamefont
  {Howl}}]{christodoulou_2022}%
  \BibitemOpen
  \bibfield  {author} {\bibinfo {author} {\bibfnamefont {M.}~\bibnamefont
  {Christodoulou}}, \bibinfo {author} {\bibfnamefont {A.}~\bibnamefont
  {Di~Biagio}}, \bibinfo {author} {\bibfnamefont {M.}~\bibnamefont
  {Aspelmeyer}}, \bibinfo {author} {\bibfnamefont {{\v{C}}.}~\bibnamefont
  {Brukner}}, \bibinfo {author} {\bibfnamefont {C.}~\bibnamefont {Rovelli}},\
  and\ \bibinfo {author} {\bibfnamefont {R.}~\bibnamefont {Howl}},\ }\bibfield
  {title} {\bibinfo {title} {{Locally Mediated Entanglement in Linearized
  Quantum Gravity}},\ }\href {https://doi.org/10.1103/PhysRevLett.130.100202}
  {\bibfield  {journal} {\bibinfo  {journal} {Phys. Rev. Lett.}\ }\textbf
  {\bibinfo {volume} {130}},\ \bibinfo {pages} {100202} (\bibinfo {year}
  {2023})},\ \Eprint {https://arxiv.org/abs/2202.03368} {arXiv:2202.03368
  [quant-ph]} \BibitemShut {NoStop}%
\bibitem [{\citenamefont {Carney}\ \emph {et~al.}(2022)\citenamefont {Carney},
  \citenamefont {Chen}, \citenamefont {Geraci}, \citenamefont {M\"uller},
  \citenamefont {Panda}, \citenamefont {Stamp},\ and\ \citenamefont
  {Taylor}}]{carney_2022}%
  \BibitemOpen
  \bibfield  {author} {\bibinfo {author} {\bibfnamefont {D.}~\bibnamefont
  {Carney}}, \bibinfo {author} {\bibfnamefont {Y.}~\bibnamefont {Chen}},
  \bibinfo {author} {\bibfnamefont {A.}~\bibnamefont {Geraci}}, \bibinfo
  {author} {\bibfnamefont {H.}~\bibnamefont {M\"uller}}, \bibinfo {author}
  {\bibfnamefont {C.~D.}\ \bibnamefont {Panda}}, \bibinfo {author}
  {\bibfnamefont {P.~C.~E.}\ \bibnamefont {Stamp}},\ and\ \bibinfo {author}
  {\bibfnamefont {J.~M.}\ \bibnamefont {Taylor}},\ }\bibfield  {title}
  {\bibinfo {title} {{Snowmass 2021 White Paper: Tabletop experiments for
  infrared quantum gravity}},\ }in\ \href@noop {} {\emph {\bibinfo {booktitle}
  {{2022 Snowmass Summer Study}}}}\ (\bibinfo {year} {2022})\ \Eprint
  {https://arxiv.org/abs/2203.11846} {arXiv:2203.11846 [gr-qc]} \BibitemShut
  {NoStop}%
\bibitem [{\citenamefont {Feng}\ and\ \citenamefont
  {Vedral}(2022)}]{Feng_2022}%
  \BibitemOpen
  \bibfield  {author} {\bibinfo {author} {\bibfnamefont {T.}~\bibnamefont
  {Feng}}\ and\ \bibinfo {author} {\bibfnamefont {V.}~\bibnamefont {Vedral}},\
  }\bibfield  {title} {\bibinfo {title} {{Amplification of gravitationally
  induced entanglement}},\ }\href {https://doi.org/10.1103/PhysRevD.106.066013}
  {\bibfield  {journal} {\bibinfo  {journal} {Phys. Rev. D}\ }\textbf {\bibinfo
  {volume} {106}},\ \bibinfo {pages} {066013} (\bibinfo {year} {2022})},\
  \Eprint {https://arxiv.org/abs/2202.09737} {arXiv:2202.09737 [quant-ph]}
  \BibitemShut {NoStop}%
\bibitem [{\citenamefont {Zhou}\ \emph {et~al.}(2022)\citenamefont {Zhou},
  \citenamefont {Marshman}, \citenamefont {Bose},\ and\ \citenamefont
  {Mazumdar}}]{Zhou_2022}%
  \BibitemOpen
  \bibfield  {author} {\bibinfo {author} {\bibfnamefont {R.}~\bibnamefont
  {Zhou}}, \bibinfo {author} {\bibfnamefont {R.~J.}\ \bibnamefont {Marshman}},
  \bibinfo {author} {\bibfnamefont {S.}~\bibnamefont {Bose}},\ and\ \bibinfo
  {author} {\bibfnamefont {A.}~\bibnamefont {Mazumdar}},\ }\bibfield  {title}
  {\bibinfo {title} {{Catapulting towards massive and large spatial quantum
  superposition}},\ }\href {https://doi.org/10.1103/PhysRevResearch.4.043157}
  {\bibfield  {journal} {\bibinfo  {journal} {Phys. Rev. Res.}\ }\textbf
  {\bibinfo {volume} {4}},\ \bibinfo {pages} {043157} (\bibinfo {year}
  {2022})},\ \Eprint {https://arxiv.org/abs/2206.04088} {arXiv:2206.04088
  [quant-ph]} \BibitemShut {NoStop}%
\bibitem [{\citenamefont {Overstreet}\ \emph {et~al.}(2023)\citenamefont
  {Overstreet}, \citenamefont {Curti}, \citenamefont {Kim}, \citenamefont
  {Asenbaum}, \citenamefont {Kasevich},\ and\ \citenamefont
  {Giacomini}}]{Overstreet_2022}%
  \BibitemOpen
  \bibfield  {author} {\bibinfo {author} {\bibfnamefont {C.}~\bibnamefont
  {Overstreet}}, \bibinfo {author} {\bibfnamefont {J.}~\bibnamefont {Curti}},
  \bibinfo {author} {\bibfnamefont {M.}~\bibnamefont {Kim}}, \bibinfo {author}
  {\bibfnamefont {P.}~\bibnamefont {Asenbaum}}, \bibinfo {author}
  {\bibfnamefont {M.~A.}\ \bibnamefont {Kasevich}},\ and\ \bibinfo {author}
  {\bibfnamefont {F.}~\bibnamefont {Giacomini}},\ }\bibfield  {title} {\bibinfo
  {title} {{Inference of gravitational field superposition from quantum
  measurements}},\ }\href {https://doi.org/10.1103/PhysRevD.108.084038}
  {\bibfield  {journal} {\bibinfo  {journal} {Phys. Rev. D}\ }\textbf {\bibinfo
  {volume} {108}},\ \bibinfo {pages} {084038} (\bibinfo {year} {2023})},\
  \Eprint {https://arxiv.org/abs/2209.02214} {arXiv:2209.02214 [quant-ph]}
  \BibitemShut {NoStop}%
\bibitem [{\citenamefont {Unruh}(2000)}]{Unruh_2000}%
  \BibitemOpen
  \bibfield  {author} {\bibinfo {author} {\bibfnamefont {W.~G.}\ \bibnamefont
  {Unruh}},\ }\bibfield  {title} {\bibinfo {title} {False loss of coherence},\
  }in\ \href@noop {} {\emph {\bibinfo {booktitle} {Relativistic Quantum
  Measurement and Decoherence}}},\ \bibinfo {editor} {edited by\ \bibinfo
  {editor} {\bibfnamefont {H.-P.}\ \bibnamefont {Breuer}}\ and\ \bibinfo
  {editor} {\bibfnamefont {F.}~\bibnamefont {Petruccione}}}\ (\bibinfo
  {publisher} {Springer Berlin Heidelberg},\ \bibinfo {address} {Berlin,
  Heidelberg},\ \bibinfo {year} {2000})\ pp.\ \bibinfo {pages}
  {125--140}\BibitemShut {NoStop}%
\bibitem [{\citenamefont {{Danielson}}\ \emph {et~al.}(2022)\citenamefont
  {{Danielson}}, \citenamefont {{Satishchandran}},\ and\ \citenamefont
  {{Wald}}}]{2022IJMPD..3141003D}%
  \BibitemOpen
  \bibfield  {author} {\bibinfo {author} {\bibfnamefont {D.~L.}\ \bibnamefont
  {{Danielson}}}, \bibinfo {author} {\bibfnamefont {G.}~\bibnamefont
  {{Satishchandran}}},\ and\ \bibinfo {author} {\bibfnamefont {R.~M.}\
  \bibnamefont {{Wald}}},\ }\bibfield  {title} {\bibinfo {title} {{Black holes
  decohere quantum superpositions}},\ }\href
  {https://doi.org/10.1142/S0218271822410036} {\bibfield  {journal} {\bibinfo
  {journal} {International Journal of Modern Physics D}\ }\textbf {\bibinfo
  {volume} {31}},\ \bibinfo {eid} {2241003} (\bibinfo {year} {2022})},\ \Eprint
  {https://arxiv.org/abs/2205.06279} {arXiv:2205.06279 [hep-th]} \BibitemShut
  {NoStop}%
\bibitem [{\citenamefont {{Danielson}}\ \emph {et~al.}(2023)\citenamefont
  {{Danielson}}, \citenamefont {{Satishchandran}},\ and\ \citenamefont
  {{Wald}}}]{Danielson2022sga}%
  \BibitemOpen
  \bibfield  {author} {\bibinfo {author} {\bibfnamefont {D.~L.}\ \bibnamefont
  {{Danielson}}}, \bibinfo {author} {\bibfnamefont {G.}~\bibnamefont
  {{Satishchandran}}},\ and\ \bibinfo {author} {\bibfnamefont {R.~M.}\
  \bibnamefont {{Wald}}},\ }\bibfield  {title} {\bibinfo {title} {{Killing
  horizons decohere quantum superpositions}},\ }\href
  {https://doi.org/10.1103/PhysRevD.108.025007} {\bibfield  {journal} {\bibinfo
   {journal} {\prd}\ }\textbf {\bibinfo {volume} {108}},\ \bibinfo {eid}
  {025007} (\bibinfo {year} {2023})},\ \Eprint
  {https://arxiv.org/abs/2301.00026} {arXiv:2301.00026 [hep-th]} \BibitemShut
  {NoStop}%
\bibitem [{\citenamefont {Gralla}\ and\ \citenamefont
  {Wei}(2024)}]{Gralla:2023oya}%
  \BibitemOpen
  \bibfield  {author} {\bibinfo {author} {\bibfnamefont {S.~E.}\ \bibnamefont
  {Gralla}}\ and\ \bibinfo {author} {\bibfnamefont {H.}~\bibnamefont {Wei}},\
  }\bibfield  {title} {\bibinfo {title} {{Decoherence from horizons: General
  formulation and rotating black holes}},\ }\href
  {https://doi.org/10.1103/PhysRevD.109.065031} {\bibfield  {journal} {\bibinfo
   {journal} {Phys. Rev. D}\ }\textbf {\bibinfo {volume} {109}},\ \bibinfo
  {pages} {065031} (\bibinfo {year} {2024})},\ \Eprint
  {https://arxiv.org/abs/2311.11461} {arXiv:2311.11461 [hep-th]} \BibitemShut
  {NoStop}%
\bibitem [{\citenamefont {Li}(2025)}]{Li_2024}%
  \BibitemOpen
  \bibfield  {author} {\bibinfo {author} {\bibfnamefont {R.}~\bibnamefont
  {Li}},\ }\bibfield  {title} {\bibinfo {title} {{Decoherence of quantum
  superpositions by Reissner-Nordstr{\"o}m black holes}},\ }\href
  {https://doi.org/10.1103/PhysRevD.111.024040} {\bibfield  {journal} {\bibinfo
   {journal} {Phys. Rev. D}\ }\textbf {\bibinfo {volume} {111}},\ \bibinfo
  {pages} {024040} (\bibinfo {year} {2025})},\ \Eprint
  {https://arxiv.org/abs/2411.04734} {arXiv:2411.04734 [hep-th]} \BibitemShut
  {NoStop}%
\bibitem [{\citenamefont {Danielson}\ \emph {et~al.}(2025)\citenamefont
  {Danielson}, \citenamefont {Satishchandran},\ and\ \citenamefont
  {Wald}}]{dswLocal_2024}%
  \BibitemOpen
  \bibfield  {author} {\bibinfo {author} {\bibfnamefont {D.~L.}\ \bibnamefont
  {Danielson}}, \bibinfo {author} {\bibfnamefont {G.}~\bibnamefont
  {Satishchandran}},\ and\ \bibinfo {author} {\bibfnamefont {R.~M.}\
  \bibnamefont {Wald}},\ }\bibfield  {title} {\bibinfo {title} {{Local
  description of decoherence of quantum superpositions by black holes and other
  bodies}},\ }\href {https://doi.org/10.1103/PhysRevD.111.025014} {\bibfield
  {journal} {\bibinfo  {journal} {Phys. Rev. D}\ }\textbf {\bibinfo {volume}
  {111}},\ \bibinfo {pages} {025014} (\bibinfo {year} {2025})},\ \Eprint
  {https://arxiv.org/abs/2407.02567} {arXiv:2407.02567 [hep-th]} \BibitemShut
  {NoStop}%
\bibitem [{\citenamefont {Hawking}\ \emph {et~al.}(2016)\citenamefont
  {Hawking}, \citenamefont {Perry},\ and\ \citenamefont
  {Strominger}}]{Hawking:2016msc}%
  \BibitemOpen
  \bibfield  {author} {\bibinfo {author} {\bibfnamefont {S.~W.}\ \bibnamefont
  {Hawking}}, \bibinfo {author} {\bibfnamefont {M.~J.}\ \bibnamefont {Perry}},\
  and\ \bibinfo {author} {\bibfnamefont {A.}~\bibnamefont {Strominger}},\
  }\bibfield  {title} {\bibinfo {title} {{Soft Hair on Black Holes}},\ }\href
  {https://doi.org/10.1103/PhysRevLett.116.231301} {\bibfield  {journal}
  {\bibinfo  {journal} {Phys. Rev. Lett.}\ }\textbf {\bibinfo {volume} {116}},\
  \bibinfo {pages} {231301} (\bibinfo {year} {2016})},\ \Eprint
  {https://arxiv.org/abs/1601.00921} {arXiv:1601.00921 [hep-th]} \BibitemShut
  {NoStop}%
\bibitem [{\citenamefont {Donnay}\ \emph {et~al.}(2018)\citenamefont {Donnay},
  \citenamefont {Giribet}, \citenamefont {Gonz\'alez},\ and\ \citenamefont
  {Puhm}}]{Donnay:2018ckb}%
  \BibitemOpen
  \bibfield  {author} {\bibinfo {author} {\bibfnamefont {L.}~\bibnamefont
  {Donnay}}, \bibinfo {author} {\bibfnamefont {G.}~\bibnamefont {Giribet}},
  \bibinfo {author} {\bibfnamefont {H.~A.}\ \bibnamefont {Gonz\'alez}},\ and\
  \bibinfo {author} {\bibfnamefont {A.}~\bibnamefont {Puhm}},\ }\bibfield
  {title} {\bibinfo {title} {{Black hole memory effect}},\ }\href
  {https://doi.org/10.1103/PhysRevD.98.124016} {\bibfield  {journal} {\bibinfo
  {journal} {Phys. Rev. D}\ }\textbf {\bibinfo {volume} {98}},\ \bibinfo
  {pages} {124016} (\bibinfo {year} {2018})},\ \Eprint
  {https://arxiv.org/abs/1809.07266} {arXiv:1809.07266 [hep-th]} \BibitemShut
  {NoStop}%
\bibitem [{\citenamefont {Rahman}\ and\ \citenamefont
  {Wald}(2020)}]{Rahman:2019bmk}%
  \BibitemOpen
  \bibfield  {author} {\bibinfo {author} {\bibfnamefont {A.~A.}\ \bibnamefont
  {Rahman}}\ and\ \bibinfo {author} {\bibfnamefont {R.~M.}\ \bibnamefont
  {Wald}},\ }\bibfield  {title} {\bibinfo {title} {{Black Hole Memory}},\
  }\href {https://doi.org/10.1103/PhysRevD.101.124010} {\bibfield  {journal}
  {\bibinfo  {journal} {Phys. Rev. D}\ }\textbf {\bibinfo {volume} {101}},\
  \bibinfo {pages} {124010} (\bibinfo {year} {2020})},\ \Eprint
  {https://arxiv.org/abs/1912.12806} {arXiv:1912.12806 [gr-qc]} \BibitemShut
  {NoStop}%
\bibitem [{\citenamefont {Ashtekar}(1987)}]{asymp-quant}%
  \BibitemOpen
  \bibfield  {author} {\bibinfo {author} {\bibfnamefont {A.}~\bibnamefont
  {Ashtekar}},\ }\href@noop {} {\emph {\bibinfo {title} {{Asymptotic
  Quantization: Based On 1984 Naples Lectures}}}},\ Monographs and Textbooks in
  Physical Science\ (\bibinfo  {publisher} {{Bibliopolis}},\ \bibinfo {address}
  {Naples, Italy},\ \bibinfo {year} {1987})\BibitemShut {NoStop}%
\bibitem [{\citenamefont {Strominger}(2017)}]{Strominger:2017zoo}%
  \BibitemOpen
  \bibfield  {author} {\bibinfo {author} {\bibfnamefont {A.}~\bibnamefont
  {Strominger}},\ }\href@noop {} {\emph {\bibinfo {title} {{Lectures on the
  Infrared Structure of Gravity and Gauge Theory}}}}\ (\bibinfo {year} {2017})\
  \Eprint {https://arxiv.org/abs/1703.05448} {arXiv:1703.05448 [hep-th]}
  \BibitemShut {NoStop}%
\bibitem [{\citenamefont {Satishchandran}\ and\ \citenamefont
  {Wald}(2019)}]{Satishchandran:2019pyc}%
  \BibitemOpen
  \bibfield  {author} {\bibinfo {author} {\bibfnamefont {G.}~\bibnamefont
  {Satishchandran}}\ and\ \bibinfo {author} {\bibfnamefont {R.~M.}\
  \bibnamefont {Wald}},\ }\bibfield  {title} {\bibinfo {title} {{Asymptotic
  behavior of massless fields and the memory effect}},\ }\href
  {https://doi.org/10.1103/PhysRevD.99.084007} {\bibfield  {journal} {\bibinfo
  {journal} {Phys. Rev. D}\ }\textbf {\bibinfo {volume} {99}},\ \bibinfo
  {pages} {084007} (\bibinfo {year} {2019})},\ \Eprint
  {https://arxiv.org/abs/1901.05942} {arXiv:1901.05942 [gr-qc]} \BibitemShut
  {NoStop}%
\bibitem [{\citenamefont {Prabhu}\ \emph {et~al.}(2022)\citenamefont {Prabhu},
  \citenamefont {Satishchandran},\ and\ \citenamefont {Wald}}]{Prabhu:2022zcr}%
  \BibitemOpen
  \bibfield  {author} {\bibinfo {author} {\bibfnamefont {K.}~\bibnamefont
  {Prabhu}}, \bibinfo {author} {\bibfnamefont {G.}~\bibnamefont
  {Satishchandran}},\ and\ \bibinfo {author} {\bibfnamefont {R.~M.}\
  \bibnamefont {Wald}},\ }\bibfield  {title} {\bibinfo {title} {{Infrared
  finite scattering theory in quantum field theory and quantum gravity}},\
  }\href {https://doi.org/10.1103/PhysRevD.106.066005} {\bibfield  {journal}
  {\bibinfo  {journal} {Phys. Rev. D}\ }\textbf {\bibinfo {volume} {106}},\
  \bibinfo {pages} {066005} (\bibinfo {year} {2022})},\ \Eprint
  {https://arxiv.org/abs/2203.14334} {arXiv:2203.14334 [hep-th]} \BibitemShut
  {NoStop}%
\bibitem [{\citenamefont {Prabhu}\ and\ \citenamefont
  {Satishchandran}(2024{\natexlab{a}})}]{Prabhu:2024zwl}%
  \BibitemOpen
  \bibfield  {author} {\bibinfo {author} {\bibfnamefont {K.}~\bibnamefont
  {Prabhu}}\ and\ \bibinfo {author} {\bibfnamefont {G.}~\bibnamefont
  {Satishchandran}},\ }\bibfield  {title} {\bibinfo {title} {{Infrared finite
  scattering theory: scattering states and representations of the BMS group}},\
  }\href {https://doi.org/10.1007/JHEP08(2024)055} {\bibfield  {journal}
  {\bibinfo  {journal} {JHEP}\ }\textbf {\bibinfo {volume} {08}},\ \bibinfo
  {pages} {055}},\ \Eprint {https://arxiv.org/abs/2402.00102} {arXiv:2402.00102
  [hep-th]} \BibitemShut {NoStop}%
\bibitem [{\citenamefont {Prabhu}\ and\ \citenamefont
  {Satishchandran}(2024{\natexlab{b}})}]{Prabhu:2024lmg}%
  \BibitemOpen
  \bibfield  {author} {\bibinfo {author} {\bibfnamefont {K.}~\bibnamefont
  {Prabhu}}\ and\ \bibinfo {author} {\bibfnamefont {G.}~\bibnamefont
  {Satishchandran}},\ }\bibfield  {title} {\bibinfo {title} {{Infrared finite
  scattering theory: Amplitudes and soft theorems}},\ }\href
  {https://doi.org/10.1103/PhysRevD.110.085022} {\bibfield  {journal} {\bibinfo
   {journal} {Phys. Rev. D}\ }\textbf {\bibinfo {volume} {110}},\ \bibinfo
  {pages} {085022} (\bibinfo {year} {2024}{\natexlab{b}})},\ \Eprint
  {https://arxiv.org/abs/2402.18637} {arXiv:2402.18637 [hep-th]} \BibitemShut
  {NoStop}%
\bibitem [{\citenamefont {Wilson-Gerow}\ \emph {et~al.}(2024)\citenamefont
  {Wilson-Gerow}, \citenamefont {Dugad},\ and\ \citenamefont
  {Chen}}]{Wilson-Gerow:2024ljx}%
  \BibitemOpen
  \bibfield  {author} {\bibinfo {author} {\bibfnamefont {J.}~\bibnamefont
  {Wilson-Gerow}}, \bibinfo {author} {\bibfnamefont {A.}~\bibnamefont
  {Dugad}},\ and\ \bibinfo {author} {\bibfnamefont {Y.}~\bibnamefont {Chen}},\
  }\bibfield  {title} {\bibinfo {title} {{Decoherence by warm horizons}},\
  }\href {https://doi.org/10.1103/PhysRevD.110.045002} {\bibfield  {journal}
  {\bibinfo  {journal} {Phys. Rev. D}\ }\textbf {\bibinfo {volume} {110}},\
  \bibinfo {pages} {045002} (\bibinfo {year} {2024})},\ \Eprint
  {https://arxiv.org/abs/2405.00804} {arXiv:2405.00804 [hep-th]} \BibitemShut
  {NoStop}%
\bibitem [{\citenamefont {Biggs}\ and\ \citenamefont
  {Maldacena}(2024)}]{Biggs:2024dgp}%
  \BibitemOpen
  \bibfield  {author} {\bibinfo {author} {\bibfnamefont {A.}~\bibnamefont
  {Biggs}}\ and\ \bibinfo {author} {\bibfnamefont {J.}~\bibnamefont
  {Maldacena}},\ }\bibfield  {title} {\bibinfo {title} {{Comparing the
  decoherence effects due to black holes versus ordinary matter}},\ }\href@noop
  {} {\  (\bibinfo {year} {2024})},\ \Eprint {https://arxiv.org/abs/2405.02227}
  {arXiv:2405.02227 [hep-th]} \BibitemShut {NoStop}%
\bibitem [{\citenamefont {Uhlmann}(1976)}]{uhlmann1976transition}%
  \BibitemOpen
  \bibfield  {author} {\bibinfo {author} {\bibfnamefont {A.}~\bibnamefont
  {Uhlmann}},\ }\bibfield  {title} {\bibinfo {title} {The “transition
  probability” in the state space of a *-algebra},\ }\href@noop {} {\bibfield
   {journal} {\bibinfo  {journal} {Reports on Mathematical Physics}\ }\textbf
  {\bibinfo {volume} {9}},\ \bibinfo {pages} {273} (\bibinfo {year}
  {1976})}\BibitemShut {NoStop}%
\bibitem [{\citenamefont {Unruh}\ and\ \citenamefont
  {Wald}(1984)}]{unruh1984happens}%
  \BibitemOpen
  \bibfield  {author} {\bibinfo {author} {\bibfnamefont {W.~G.}\ \bibnamefont
  {Unruh}}\ and\ \bibinfo {author} {\bibfnamefont {R.~M.}\ \bibnamefont
  {Wald}},\ }\bibfield  {title} {\bibinfo {title} {What happens when an
  accelerating observer detects a rindler particle},\ }\href@noop {} {\bibfield
   {journal} {\bibinfo  {journal} {Physical Review D}\ }\textbf {\bibinfo
  {volume} {29}},\ \bibinfo {pages} {1047} (\bibinfo {year}
  {1984})}\BibitemShut {NoStop}%
\bibitem [{\citenamefont {Hawking}(1975)}]{hawking1975particle}%
  \BibitemOpen
  \bibfield  {author} {\bibinfo {author} {\bibfnamefont {S.~W.}\ \bibnamefont
  {Hawking}},\ }\bibfield  {title} {\bibinfo {title} {Particle creation by
  black holes},\ }\href@noop {} {\bibfield  {journal} {\bibinfo  {journal}
  {Communications in mathematical physics}\ }\textbf {\bibinfo {volume} {43}},\
  \bibinfo {pages} {199} (\bibinfo {year} {1975})}\BibitemShut {NoStop}%
\bibitem [{\citenamefont {Wald}(1975)}]{Wald_75}%
  \BibitemOpen
  \bibfield  {author} {\bibinfo {author} {\bibfnamefont {R.~M.}\ \bibnamefont
  {Wald}},\ }\bibfield  {title} {\bibinfo {title} {On particle creation by
  black holes},\ }\href {https://doi.org/10.1007/BF01609863} {\bibfield
  {journal} {\bibinfo  {journal} {Communications in Mathematical Physics}\
  }\textbf {\bibinfo {volume} {45}},\ \bibinfo {pages} {9} (\bibinfo {year}
  {1975})}\BibitemShut {NoStop}%
\bibitem [{\citenamefont {Wald}(1995)}]{Wald_1995}%
  \BibitemOpen
  \bibfield  {author} {\bibinfo {author} {\bibfnamefont {R.~M.}\ \bibnamefont
  {Wald}},\ }\href@noop {} {\emph {\bibinfo {title} {{Quantum Field Theory in
  Curved Space-Time and Black Hole Thermodynamics}}}},\ Chicago Lectures in
  Physics\ (\bibinfo  {publisher} {University of Chicago Press},\ \bibinfo
  {address} {Chicago, IL},\ \bibinfo {year} {1995})\BibitemShut {NoStop}%
\bibitem [{\citenamefont {{Danielson}}(2025)}]{IDT}%
  \BibitemOpen
  \bibfield  {author} {\bibinfo {author} {\bibfnamefont {D.~L.}\ \bibnamefont
  {{Danielson}}},\ }\bibfield  {title} {\bibinfo {title} {{Horizon Algebras and
  Soft Quantum Information}},\ }\href@noop {} {\bibfield  {journal} {\bibinfo
  {journal} {to appear}\ } (\bibinfo {year} {2025})}\BibitemShut {NoStop}%
\bibitem [{\citenamefont {{Brasco}}\ \emph {et~al.}(2020)\citenamefont
  {{Brasco}}, \citenamefont {{G{\'o}mez-Castro}},\ and\ \citenamefont
  {{V{\'a}zquez}}}]{Brasco_2020}%
  \BibitemOpen
  \bibfield  {author} {\bibinfo {author} {\bibfnamefont {L.}~\bibnamefont
  {{Brasco}}}, \bibinfo {author} {\bibfnamefont {D.}~\bibnamefont
  {{G{\'o}mez-Castro}}},\ and\ \bibinfo {author} {\bibfnamefont {J.~L.}\
  \bibnamefont {{V{\'a}zquez}}},\ }\bibfield  {title} {\bibinfo {title}
  {{Characterisation of homogeneous fractional Sobolev spaces}},\ }\href
  {https://doi.org/10.48550/arXiv.2007.08000} {\bibfield  {journal} {\bibinfo
  {journal} {arXiv e-prints}\ ,\ \bibinfo {eid} {arXiv:2007.08000}} (\bibinfo
  {year} {2020})},\ \Eprint {https://arxiv.org/abs/2007.08000}
  {arXiv:2007.08000 [math.AP]} \BibitemShut {NoStop}%
\bibitem [{\citenamefont {Kholevo}(1972)}]{kholevo1972quasiequivalence}%
  \BibitemOpen
  \bibfield  {author} {\bibinfo {author} {\bibfnamefont {A.}~\bibnamefont
  {Kholevo}},\ }\bibfield  {title} {\bibinfo {title} {On quasiequivalence of
  locally normal states},\ }\href@noop {} {\bibfield  {journal} {\bibinfo
  {journal} {Theoretical and Mathematical Physics}\ }\textbf {\bibinfo {volume}
  {13}},\ \bibinfo {pages} {1071} (\bibinfo {year} {1972})}\BibitemShut
  {NoStop}%
\bibitem [{\citenamefont {{B{\'e}ny}}\ and\ \citenamefont
  {{Oreshkov}}(2010)}]{2010PhRvL.104l0501B}%
  \BibitemOpen
  \bibfield  {author} {\bibinfo {author} {\bibfnamefont {C.}~\bibnamefont
  {{B{\'e}ny}}}\ and\ \bibinfo {author} {\bibfnamefont {O.}~\bibnamefont
  {{Oreshkov}}},\ }\bibfield  {title} {\bibinfo {title} {{General Conditions
  for Approximate Quantum Error Correction and Near-Optimal Recovery
  Channels}},\ }\href {https://doi.org/10.1103/PhysRevLett.104.120501}
  {\bibfield  {journal} {\bibinfo  {journal} {\prl}\ }\textbf {\bibinfo
  {volume} {104}},\ \bibinfo {eid} {120501} (\bibinfo {year} {2010})},\ \Eprint
  {https://arxiv.org/abs/0907.5391} {arXiv:0907.5391 [quant-ph]} \BibitemShut
  {NoStop}%
\bibitem [{\citenamefont {Kudler-Flam}\ and\ \citenamefont
  {Penington}(2025)}]{Kudler-Flam:2025yur}%
  \BibitemOpen
  \bibfield  {author} {\bibinfo {author} {\bibfnamefont {J.}~\bibnamefont
  {Kudler-Flam}}\ and\ \bibinfo {author} {\bibfnamefont {G.}~\bibnamefont
  {Penington}},\ }\bibfield  {title} {\bibinfo {title} {{It costs nothing to
  teleport information into a black hole}},\ }\href@noop {} {\  (\bibinfo
  {year} {2025})},\ \Eprint {https://arxiv.org/abs/2504.01058}
  {arXiv:2504.01058 [hep-th]} \BibitemShut {NoStop}%
\end{thebibliography}%

\end{document}